\documentclass[prb,aps,twocolumn,nofootinbib,showpacs,superscriptaddress,floatfix]{revtex4}
\usepackage{graphicx}
\usepackage{amsfonts}
\usepackage{amsbsy,bbm,amssymb,amsmath}
\usepackage{dcolumn}
\usepackage{bm}
\usepackage{epsfig}
\usepackage{color}
\newcommand{\bra}[1]{\langle#1|}

\newcommand{\ket}[1]{|#1\rangle}

\newcommand{\ketbra}[2]{|\,{#1} \, \rangle\hspace{-0.05cm} \langle \, {#2} \,|  }

\begin{document}
\title{High-fidelity two-qubit gates via dynamical decoupling of local $1/f$ noise \\ at optimal point}

\author{A. D'Arrigo}
\affiliation{Dipartimento di Fisica e Astronomia,
Universit\`a di Catania, Via Santa Sofia 64, 95123 Catania, Italy}
\affiliation{Centro Siciliano di Fisica Nucleare e Struttura della Materia,Via Santa Sofia 64, 95123 Catania, Italy}
\author{G. Falci}
\affiliation{Dipartimento di Fisica e Astronomia,
Universit\`a di Catania, Via Santa Sofia 64, 95123 Catania, Italy}
\affiliation{CNR-IMM  UOS Catania (Universit\`a), 
Consiglio Nazionale delle Ricerche, Via Santa Sofia 64, 95123 Catania, Italy}
\affiliation{Istituto Nazionale di Fisica Nucleare, Sezione di Catania, Via Santa Sofia 64, 95123 Catania, Italy}
\author{E. Paladino}
\affiliation{Dipartimento di Fisica e Astronomia,
Universit\`a di Catania, Via Santa Sofia 64, 95123 Catania, Italy}
\affiliation{CNR-IMM  UOS Catania (Universit\`a), 
Consiglio Nazionale delle Ricerche, Via Santa Sofia 64, 95123 Catania, Italy}
\affiliation{Istituto Nazionale di Fisica Nucleare, Sezione di Catania, Via Santa Sofia 64, 95123 Catania, Italy}

\begin{abstract}
We investigate the possibility to achieve high-fidelity universal two-qubit gates by supplementing optimal tuning of individual qubits with dynamical decoupling (DD) of local $1/f$ noise. We consider simultaneous local pulse sequences applied during the gate operation and  compare the efficiencies of periodic, Carr-Purcell and Uhrig DD with hard 
$\pi$-pulses along two directions ($\pi_{z/y}$ pulses). 
We present analytical perturbative results (Magnus expansion) in the quasi-static noise approximation combined with
numerical simulations for realistic $1/f$ noise spectra.
The gate efficiency is studied as a function of the gate duration, of the number $n$ of
pulses, and of the high-frequency roll-off. We find that the gate error is non-monotonic in $n$,
decreasing as $n^{-\alpha}$ in the asymptotic limit, $\alpha \geq 2$ depending on the DD sequence.
In this limit $\pi_z$-Urhig is the most efficient scheme for quasi-static $1/f$ noise, but it is highly sensitive to the soft UV-cutoff. For small number of pulses, $\pi_z$ control yields anti-Zeno behavior, whereas
$\pi_y$ pulses minimize the error for a finite $n$.
For the current noise figures in superconducting qubits,
two-qubit gate errors $\sim 10^{-6}$, meeting the requirements for fault-tolerant quantum
computation, can be achieved. The Carr-Purcell-Meiboom-Gill sequence is the
 most efficient procedure, stable for $1/f$ noise with UV-cutoff up to gigahertz.
\end{abstract}

\date{\today}


\maketitle

\section{Introduction}
Universal two-qubit gates represent an essential ingredient for digital quantum computation \cite{nielsenchuang}.
A central challenge for quantum based technologies is fighting decoherence which arises from unwanted interactions
with environmental degrees of freedom, resulting in errors both during manipulation and 
transmission of quantum information. 
High-fidelity quantum information processing and fault-tolerant quantum computing in fact set strict limits on allowable 
errors per gate \cite{Knill1998,Preskill1998,Knill2005,Aliferis2006}. 
This issue is particularly relevant for solid state nanoscale devices. Both semiconducting and superconducting
nanocircuits are hindered by the presence of material-inherent non-Markovian fluctuations often characterized by a $1/f^\alpha$
($\alpha \sim 1$) power spectrum~\cite{PaladinoRMP}. 

One successful strategy to increase phase-coherence times is to operate qubits at optimal working points, where
low-frequency noise effects vanish to the lowest order~\cite{Vion2002,ithier2005PRB}.
This concept has been generalized to more complex architectures  
and proved to be efficient in reducing defocussing due to $1/f$ noise also in two-qubit gates \cite{Paladino2010,Paladino2011}. 
Optimal tuning is a {\em passive} (or error avoiding) stabilization strategy which, in combination with improved materials
and environment engineering, has led to high-fidelity superconducting~\cite{PaladinoRMP} and semiconducting
quantum dot charge- \cite{Petersson2010} and spin-qubits 
~\cite{Taylor2013,Medford2013,Fei2015,Martins2016,Reed2016}.
Dynamical decoupling (DD)  is an {\em active} (or error correcting) scheme,  based 
on the repeated application of control pulses designed to coherently average out unwanted  interactions with the environment~\cite{viola1998PRA,viola1999PRL}. 
Improvement against $1/f$ noise in single-qubit gates via DD \cite{falci2004PRA,faoro2004PRL,Gutmann2005,bergli2007PRB,cywinsky2008PRB,lutchyn2008PRB,Cheng2008,Rebentrost2009},
composite pulses \cite{Mottonen2006,Chen2012,Wang2014,Kabytayev2014}, dynamically corrected gates \cite{Khodjasteh2012,Kabytayev2014} 
and quantum optimal control \cite{Montangero2007,Whaley2012}, has been demonstrated.
In superconducting qubits pulsed control has been exploited to reduce dephasing
due to charge- and magnetic flux-noise~\cite{nakamura2002PRL,Bertet2005,ithier2005PRB,Yoshihara2006,bylander2011NatPhys,yan2012PRB,Yuge2011PRL,wang2007PRL,roy2011PRA}.
Recently DD of pure dephasing due to quadratic coupling to Gaussian distributed $1/f^\alpha$ noise  
has been used as a tool for noise spectroscopy\cite{Cywinski2014}.

The considerable achievements reached by dynamical decoupling of non-Markovian noise in single qubit
gates, has stimulated increasing interest in extending these techniques to multi-qubit systems, i.e. registers and
gates\cite{Souza2015,Huang2014,West2010PRL,Piltz2013,Souza2012,Khodjasteh2009a,Khodjasteh2009b,Khodjasteh2010,Ng2011}. Recently, the lifetime of an entangled state of a superconducting flux qubit coupled to a microscopic two-level 
fluctuator has been enhanced by DD techniques \cite{gustavsson2012PRL}.  
Entanglement preservation between independent qubits (storage) from local pure dephasing random-telegraph and $1/f$ noise 
by dynamical decoupling has been predicted \cite{Darrigo2014Annals,LoFranco2014}  
and demonstrated in an all-optical experiment \cite{Orieux2015}. 

Integration of DD with quantum {\em gates} is still a non trivial challenge.
Achieving sufficiently high-fidelity two-qubit {\em gates}, compatible with fault-tolerance requirements,
requires satisfying several constraints: decoupling should not alter the gate operation, it should not
make the gate last too long in order to avoid decoherence from Markovian channels, it should not
increase the computational complexity limiting scalability. Moreover it is desirable that each individual
qubit operates at its own optimal working point. 

In this paper we investigate the possibility of achieving high-fidelity  $\sqrt{\rm{i-SWAP}}$ gate 
by dynamical decoupling of local  $1/f$ noise. 
We assume that both qubits operate at their optimal working point where 
noise is transverse with respect to the single qubit Hamiltonians (see Eq. (\ref{eq:Hnoise})).  
We investigate the gate efficiency under the application of local DD sequences designed to be ineffective in the absence of coupling to noise sources. Pulses are applied during the gate operation, in a decouple while compute strategy. 
We consider DD sequences currently available in the labs and conveniently exploited with superconducting qubits \cite{bylander2011NatPhys,yan2012PRB,Yuge2011PRL,wang2007PRL,roy2011PRA}, namely the
periodic DD (PDD) \cite{viola1998PRA,viola1999PRL}, Carr-Purcell (CP) \cite{carr1954PhysRev}, Carr-Purcell-Meiboom-Gill (CPMG)  \cite{meiboom1958RevSci} and Uhrig DD (UDD) sequences \cite{uhrig2007PRL}.
This choice is attractive since it is based on standard, simple sequences already experimentally available.
Our strategy does not require additional calibration efforts besides single qubit optimal tuning, it does not limit scalability since it acts locally, it does not involve encoding in a larger Hilbert space and it does not  make the gate lasting longer, as in some DD schemes. Our analysis points out
achievable advantages and at the same time possible constraints of the simplest decoupling schemes currently
available. More complex sequences, for instance concatenated, recursive etc. tailored to the specific two-qubit setting, may allow for further optimization.        

We compare the efficiencies of PDD, CP and UDD by applying hard $\pi$-pulses along two directions 
($\hat z $ and $\hat y$). The corresponding qubit rotations around two axis of the Bloch sphere ($\pi_{z/y}$ pulses) coherently average out the transverse noise interaction term.    
Our analysis is based on the exact numerical solution of the stochastic Sch\"odinger equation  (SSE) for the coupled-qubits 
evolution in the presence of classical noise with $1/f$ power spectrum.  
For the PDD the gate error is also evaluated analytically making a perturbative (Magnus) expansion of the propagator for quasi-static noise
\cite{falci2005PRL,ithier2005PRB}. In our analysis we consider realistic $1/f$ spectra as detected in superconducting nanodevices \cite{Vion2002,bylander2011NatPhys,yan2012PRB}   
and discuss the stability of the procedure with increasing high-frequency (soft) cut-off of the spectrum.

We find a non-monotonic behavior of the gate error with the number of applied pulses, $n$.
In the asymptotic limit of $n$ much larger than a threshold $n_0$, which depends on the qubits coupling strength,  
the gate error decreases as $n^{-\alpha}$, with a sequence specific $\alpha \geq 2$.
In this regime $\pi_z$-UDD is the most efficient sequence against quasi-static $1/f$ noise, but 
it is highly sensitive to the soft UV-cutoff. 
For $n \leq n_0$, operating with $\pi_z$ pulses increases the error with respect
to the unconditioned evolution, a behavior reminiscent of the anti-Zeno effect in single qubit gates \cite{falci2004PRA,falci2005PhysE,bergli2007PRB,paladino2003ASSP}. For $\pi_y$ pulses instead 
the error shows a minimum as a function of the number of applied pulses at  $n \lesssim n_0$.
For typical noise figures in superconducting qubits, the most
efficient error reduction, stable for $1/f$ noise with UV-cutoff up to gigahertz, 
is achieved by the CPMG sequence with about ten pulses. 

In Section II the model, the considered DD sequences and the methods are illustrated. In Section III we present our
results for the gate error and discuss its dependence on relevant physical parameters (gate duration, spectrum bandwidth and amplitude). In Section IV we comment our findings and draw our conclusions in Section V.

\section{Model and dynamical decoupling sequences}

We consider  two resonant qubits with a coupling term transverse with respect to 
the qubits quantization axis, as modeled by
\begin{equation}
\mathcal{H}_0 = 
 -\frac{\Omega}{2} \, \sigma_{1z} \otimes \mathbb{I}_{2}
 -\frac{\Omega}{2}  \, \mathbb{I}_{1} \otimes \sigma_{2z}
+ \frac{\omega_c}{2} \, \sigma_{1x} \otimes \sigma_{2x}  \, .
\label{H0}
\end{equation}
Here $\sigma_{\alpha z}$ are Pauli matrices, $\mathbb{I}_{\alpha}$ is the identity
in  qubit-$\alpha$ Hilbert space ($\alpha=1,2$), we put $\hbar=1$.
This model applies in particular to the fixed, capacitive or inductive, coupling of
superconducting qubits~\cite{Clarke2008,PaladinoRMP} 
individual-qubit control allows an effective switch on/off of the interaction. 
Eigenvalues and eigenvectors of Eq.~(\ref{H0})  are reported in  Table \ref{tab:eigensystem0}.
\begin{table}[t!]
{\begin{tabular}{ccc}
\toprule
$i$ & $\omega_i$ & $| i \rangle$ \\
\colrule
0 & $- \sqrt{\Omega^2 + (\omega_c/2)^2}$ & 
$- ( \sin \varphi/2  )| ++ \rangle+ ( \cos \varphi/2 ) | -- \rangle$ \\
1 & $-\omega_c/2$ & $(- | +- \rangle + | -+ \rangle )/\sqrt{2}$ \\
2 & $\omega_c/2$ & $( | +- \rangle + | -+ \rangle )/\sqrt{2}$ \\
3 & $\sqrt{\Omega^2 + (\omega_c/2)^2}$ & 
$( \cos \varphi/2 )| ++ \rangle + ( \sin \varphi/2) 
| -- \rangle $ \\
\botrule
\end{tabular}}
\caption{\label{tab:eigensystem0} \footnotesize
Eigenvalues and eigenvectors of ${\mathcal{H}}_0 $ expressed in the computational
basis  $| \mu \nu \rangle \equiv | \mu \rangle_1 \otimes
| \nu \rangle_2$,  $\mu,\nu \in \{ +,- \}$ with
$\sigma_{\alpha z} | \pm \rangle_\alpha =
\mp | \pm \rangle_\alpha$ and $\tan \varphi = - \omega_c/(2 \Omega)$.}
\end{table} 
If the two qubits are prepared in the factorized state $| +- \rangle$, free evolution
under Eq.~(\ref{H0}) for a time $t_e= \pi /2 \omega_c$  realizes a $\sqrt{{\rm i-SWAP}}$ operation
$| \psi(t_e) \rangle = [| +- \rangle - i | -+ \rangle]/\sqrt 2 \equiv \ket{\psi_e}$.
The dynamics takes place inside the $\{ |1 \rangle, |2 \rangle\}$ subspace, which we name "SWAP-subspace". 

We suppose that each qubit operates at its optimal point, where noise is transverse with respect
the single qubit Hamiltonian
\begin{equation}
\label{eq:Hnoise}
\delta \mathcal{H}(t) = -\frac{1}{2} x_{1}(t) \, \sigma_{1x} \otimes \mathbb{I}_{2}
- \frac{1}{2} \mathbb{I}_{1} \otimes  x_{2}(t) \, \sigma_{2x}  \,,
\end{equation}
here $x_\alpha(t) $ is a stochastic process whose power spectrum
\begin{equation}
S_\alpha(\omega) = \int_{-\infty }^\infty d t \, \langle x_\alpha(t) x_\alpha(0) \rangle \, e^{i \omega t}
\label{def_spectrum}
\end{equation}
is $1/f$ in the frequency  range $f \in [\gamma_{m,\alpha}, \gamma_{M,\alpha}]$. In the following, for the
sake of simplicity, we will assume identical local noise characteristics, $S_\alpha(\omega) \equiv S(\omega)$
and frequency range, $\gamma_{m,\alpha}=\gamma_m$,  $\gamma_{M,\alpha}=\gamma_M$.

We consider  dynamical decoupling protocols consisting of 
instantaneous pulses acting locally and simultaneously on each qubit. The system Hamiltonian under DD takes the form
\begin{equation}
\tilde{\cal H}(t)\,=\, 
{\cal H}_0+\delta{\cal H}(t) +{\cal V}(t)
\label{eq:controlled-semiclassical-noisyH}
\end{equation}
where
${\cal V}(t)\,=\, {\cal V}_1(t)\otimes\mathbb{I}_{2}\, + \,\mathbb{I}_{1}\otimes{\cal V}_2(t)$,
and ${\cal V}_\alpha(t)$ denotes the action of a sequence of local operations on qubit $\alpha$ 
applied at times $t=t_i$, $i \in \{1,m\}$.
The control sequence is designed to reduce the effect of noise acting along $\sigma_{\alpha x}$ without
altering the gate operation.  
These two requirements are fulfilled applying an {\em even number} of simultaneous $\pi$-pulses around 
the $z$ or the $y$-axis of the Bloch sphere of each qubit, denoted respectively as $\pi_z$, $\pi_y$.

The pulses are applied at times $t_i=\delta_i t_e$, where $0\leq\delta_i\leq1$ with $i=1,\ldots,m$. 
For the PDD sequence $\delta_i=i/m$, with the last pulse applied at time $t_e$, the pulse interval being $\Delta t=t_e/m$. 
A PDD sequence with $m=2$ corresponds to the echo procedure. 
For the CP ($\pi_z$ pulses) and the CPMG ($\pi_y$ pulses) sequences it is $\delta_i =(i-1/2)/m$.
For the UDD  sequence $\delta_i=\sin^2[\pi i/(2m+2)]$. 
In the limit of a two-pulse cycle, $m=2$, UDD reduces to the CP sequence.

The system density matrix $\rho(t)$ can be expressed as
\begin{equation}
\rho(t) = \int {\mathcal D}[{\bf x}(t)] \,  P[{\bf x}(t)] \;  \rho\big(t | {\bf x}(t)\big) \, 
\end{equation}
where ${\bf x}(t)=\{x_1(t),x_2(t)\}$, and  $\rho\big(t | {\bf x}(t)\big)$ is the density matrix 
for a  given realization, ${\bf x}(t)$, of the stochastic process whose probability density function is $P[{\bf x}(t)]$.
For a sequence of two pulses separated by $\Delta t$ and a given realization
${\bf x}(t)$, the propagator reads (with $t_0=0$)
\begin{eqnarray}
    && U(t_{i+1},t_{i-1}|{\bf x}(t)) = \nonumber \\
&&={\cal S}\,\hat{T}e^{-i\int_{t_i}^{t_{i+1}}{\cal H}(t')dt'}\,{\cal S}\,
    \hat{T} e^{-i\int_{t_{i-1}}^{t_{i}}{\cal H}(t')dt'} \, ,
\label{eq:-a1}
\end{eqnarray}
where ${\cal S} = {\cal S}_1 \otimes {\cal S}_2$ denotes the pulse propagator factorized in local actions. 
It is easily seen that two pulses separated by $\Delta t$ eliminate the leading order 
effect of noise provided $\Delta t \ll 1/ \gamma_M$.
In fact, under this condition it is possible to perform a quasi-static approximation \cite{falci2005PRL,ithier2005PRB} 
replacing ${\bf x}(t)$ with a stochastic static value ${\bf x}$. 
The first order expansion in $\Delta t$ of Eq.(\ref{eq:-a1}) reads
\begin{eqnarray}
U(t_{i+1} \!\!&,& \!\!t_{i-1}|{\bf x}(t)) \simeq \nonumber \\  
&& \simeq {\cal S}\,\Big(\mathbb{I}-i{\cal H}(t_{i})\Delta t\Big)\,{\cal S}\, 
\Big(\mathbb{I}-i{\cal H}(t_{i-1})\Delta t\Big)  
\nonumber \\
&&\simeq  \mathbb{I}  - i \, \mathcal{\tilde H} 2 \Delta t \simeq e^{ -  i \, \mathcal{\tilde H} 2 \Delta t }\, ,
\label{eq:-a2}
\end{eqnarray}
where ${\mathcal{\tilde H}} = ({\cal S} {\cal H} {\cal S} + {\cal H})/2 $. For $\pi_z$ pulses $\mathcal{\tilde H} = \mathcal{H}_0$, whereas for $\pi_y$ pulses 
$\mathcal{\tilde H} = (\omega_c/2) \sigma_{1x} \otimes \sigma_{2x}$.
Thus, the sequence of the simultaneous $\pi_z$ pulses on the two qubits 
cancels, to the first order in $\Delta t$, the effect of any noise realization on the coupled qubits provided 
pulses are sufficiently frequent. The sequences of $\pi_y$ pulses also cancels the single qubit dynamics.
In both cases, the action of the approximated propagator on the factorized state $\ket{+-}= [\ket{1}+\ket{2}]/\sqrt{2}$
corresponds to the $\sqrt{\rm{i-SWAP}}$ operation at time  $2 \Delta t = t_e$.
This result is also valid for a sequence of an even number 
of pulses provided that $t_e \ll 1/\gamma_M$.

\section{Dynamical decoupling of transverse $1/f$ noise}

In order to estimate the gate performance under DD  we evaluate the fidelity with respect to the desired target state 
$| \psi_e \rangle$,  ${\cal F}$, and the corresponding error $\varepsilon$ defined as
\begin{equation}
{\cal F}\,=\,\bra{\psi_e}\rho(t_e)\ket{\psi_e}, \qquad \varepsilon\,=\,1-\,{\cal F}
\label{eq:gate-fidelity-error}
\end{equation}
where 
\begin{equation}
\bra{\psi_e}\rho(t_e)\ket{\psi_e}  = \int {\mathcal D}[{\bf x}(t)] \,  P[{\bf x}(t)] \; 
\bra{\psi_e} \rho\big(t | {\bf x}(t)\big)  \ket{\psi_e} \, .
\end{equation}
In the following these quantities are obtained by the exact numerical solution of the stochastic Schr\"odinger 
equation (SSE) for the coupled-qubits evolution under the action of the considered dynamical decoupling sequences.
Noise with $1/f$ spectrum is simulated as the superposition of random telegraph
processes with switching rates $\gamma$ distributed as $1/\gamma$ in $[\gamma_m,\gamma_M]$ \cite{weissman1988RMP,PaladinoRMP}. 
The spectrum reads $S(\omega)\approx {\mathcal A}/ \omega$, with $\mathcal A = \pi\Sigma^2/\ln(\gamma_M/\gamma_m))$, for $\omega \leq \gamma_M/2 \pi$, 
with a roll-off to $1/f^2$ behavior at higher frequencies; $\Sigma^2$ is the noise variance.
The number of noise realizations over which the average is performed is quite large ($N \geq 10^4$ unless specified). Under
this condition, the numeric simulation can be considered a reliable method for calculating the gate error.
As we shall see, for realistic noise figures, the quasi-static approximation ${\bf x}(t) \approx {\bf x}$ (numerical
or analytical) captures the system's evolution for times of interest, $t \leq 1/\gamma_M$.    

As a case study we consider  persistent-current flux qubits and
magnetic flux noise characteristics reported in the experiments \cite{bylander2011NatPhys,yan2012PRB}
where $1/f$-type power laws in the frequency ranges $0.2$-$20$ MHz and $0.01$-$100$ Hz has been detected. 
Based on these results, we assume a flux noise $S_\Phi (\omega) = A_\Phi/ (2 \pi \omega)$ extending between $1$~Hz and 
$10$~MHz with amplitude $A_\Phi=(1.7 \times 10^{-6} \, \Phi_0)^2$ and $\gamma \in[1, 10^7]$~Hz ($\Phi_0$ is the magnetic flux quantum) and noise variance
$\Sigma \approx 2\pi\times10^7$ Hz ($\Omega \approx 2 \pi \times 5$GHz). 
For comparison we will also consider charge-phase qubits and charge noise 
whose variance is $\Sigma_x \approx 2 \times 10^{-2} \,\Omega $ 
in $\gamma_m \approx 1$Hz, $\gamma_M \approx 1$MHz ($\Omega \approx 2 \pi  \times 16$GHz)  \cite{Vion2002}.

\subsection{Dynamical Decoupling with $\pi_z$ pulses}

\subsubsection{Periodic Dynamical Decoupling}
We start our analysis considering the PDD sequence. 
For this sequence the gate error can be easily evaluated analytically 
making a perturbative (Magnus) expansion the propagator assuming quasi-static noise \cite{ithier2005PRB,falci2005PRL}.
For a PPD sequence with pulses applied at times $t_k=k\, t_e /(2n)$, $k=1, \dots 2n$,
we suppose that $t_e = 2 n \Delta t \ll 1/\gamma_M$.
Replacing ${\bf x}(t)$ with static random ${\bf x}$ during the entire evolution time $t_e$, we obtain 
\begin{eqnarray}
\bra{\psi_e} \rho\big(t | {\bf x}(t)\big)  \ket{\psi_e}  
&\approx &\bra{\psi_e} \mathcal U (t | {\bf x}) \rho(0)  \mathcal U^\dagger (t |{\bf x})  \ket{\psi_e} 
\nonumber \\ 
&= & |\bra{\psi_e} \mathcal U(t,{\bf x}) \ket{+-}|^2
\end{eqnarray}
where we used $\rho(0)=  \ketbra{+-}{+-}$ and
\begin{eqnarray}
\mathcal U(t_e | {\bf x}) &=& ( \mathcal S e^{- i \mathcal H_{qs}\Delta t} 
 \mathcal S e^{- i \mathcal H_{qs}\Delta t} )^n  \nonumber \\
&=& ( e^{ - i \tilde{\mathcal H}_{qs} \Delta t} e^{- i \mathcal H_{qs}\Delta t}  )^n \, ,
\label{eq:propagator}
\end{eqnarray}
here $\mathcal H_{qs}$ is the quasi-static Hamiltonian given by Eqs. (\ref{H0}),(\ref{eq:Hnoise}) with $x_{\alpha}(t)$ replaced by
$x_\alpha$, the fixed value during the noise realization and 
$\tilde{\mathcal H}_{qs} =  \mathcal H_0 - \delta \mathcal H$. 

We evaluate the propagator Eq.(\ref{eq:propagator}) by the Magnus expansion up to the third order.
We obtain
\begin{eqnarray}
\mathcal U(t_e, {\bf x}) & \approx & e^{- i \mathcal H_{eff}(\Delta t) t_e} \\
\mathcal H_{eff}(\Delta t) &=& \mathcal H_0 + \frac{\Omega \Delta t}{2} \, 
x_1 (-\sigma_{1y} + \frac{2}{3} x_1 \Delta t \sigma_{1z}) \otimes \mathbb{I}_2 
\nonumber \\
&+& \frac{\Omega \Delta t}{2} \,  \mathbb{I}_1 \otimes 
 x_2 (-\sigma_{2y} + \frac{2}{3} x_2 \Delta t \sigma_{2z}) 
\, ,
\label{Magnus-3}
\end{eqnarray}
where we used the condition $t_e= 2 n \Delta t$. 
Evaluating the fourth order term of the Magnus expansion at time $t_e$ we find that Eq.(\ref{Magnus-3}) is valid 
for $\Delta t$ sufficiently small, or equivalently if
the number of pulses is larger than $n_0= (\pi / 8 \sqrt 3) \Omega/\omega_c$.
We observe that the extra terms in the effective Hamiltonian consist of linear 
transverse contributions (terms $x_i \sigma_{iy}$) 
which are $\propto \Delta t$ and of quadratic longitudinal contributions (terms $x_i^2 \sigma_{iz}$) 
scaling as $\Delta t^2$. Both terms can be treated in perturbation theory.
The gate error is of the second order in $\Delta t$ and, for fixed realization of the stochastic processes $x_i$, reads
\begin{eqnarray}
\varepsilon &=& {\Delta t^2 \over 8} (x_1^2+x_2^2)\Big [1- {1 \over \sqrt 2} \cos \Big (\sqrt{\Omega^2 + {\omega_c^2 \over 4}}t_e \Big ) \nonumber \\
&-& {1 \over 2 \sqrt 2} {\omega_c \over \sqrt{\Omega^2 + \omega_c^2/4}} \sin \Big ( \sqrt{\Omega^2 + {\omega_c^2 \over 4}}t_e \Big ) \Big ]
\label{error}
\end{eqnarray}
\begin{figure}[t!]
\begin{center}
{
\includegraphics[width=0.40\textwidth]{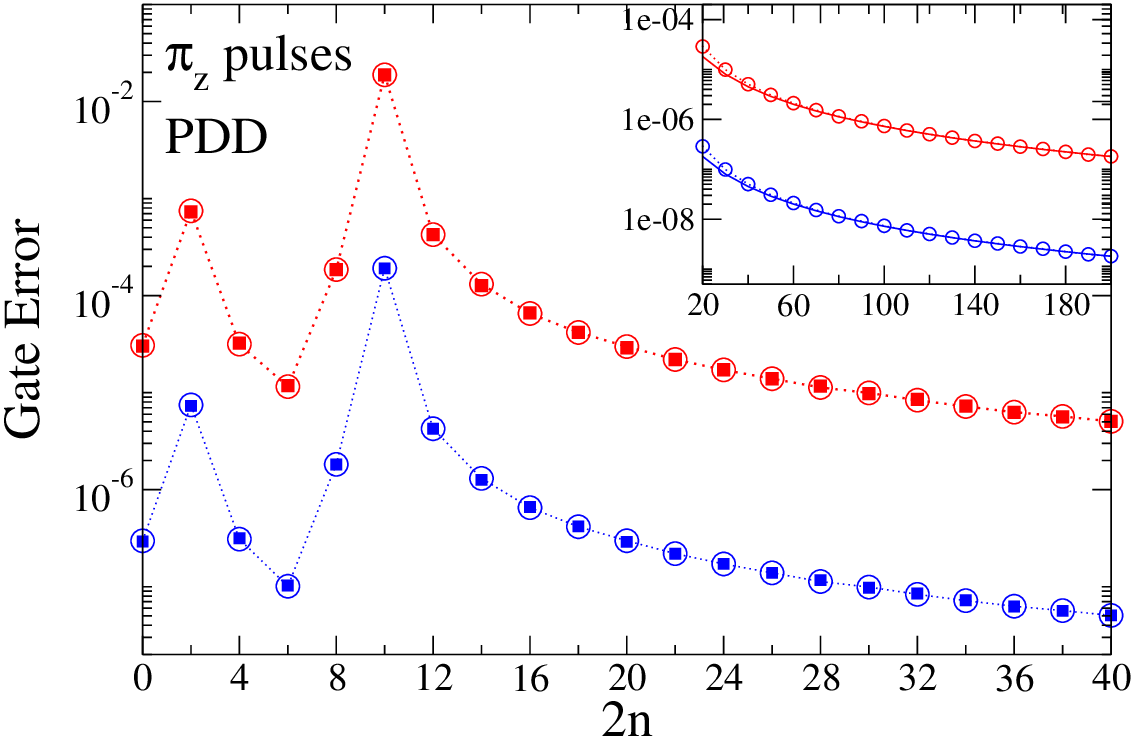} 
}
\caption{\label{figure1}\footnotesize (Color online)
Gate error under PDD at time $t_e= \pi/(2\omega_c)= 2n \Delta t$ ($n \geq 1$) as a function of 
the number of couple of pulses. 
The point at $n=0$ represents the gate error in the absence of DD. 
The lower curve (blue) corresponds to 
$\Sigma_1=\Sigma_2= 10^8$~rad/s, upper curve (red) corresponds to $\Sigma_1=\Sigma_2=  10^9$~rad/s, 
in both cases $\gamma_m= 1$s$^{-1}$, $\gamma_M= 10^6$ s$^{-1}$. 
Squares are the errors in the quasi-static approximation 
(numerical), empty circles are the exact result from the solution of the SSE.
Inset: for $n \gg n_0 \approx 5$, the error decays as $n^{-2}$, as given by 
Eq.(\ref{av-errorPDD}) (continuous lines). 
Parameters are $\omega_c= 5 \times 10^9$~rad/s, $\Omega= 10^{11}$rad/s.
}
\end{center}
\end{figure}
\begin{figure*}[t!]
  \begin{center}
\includegraphics[width=0.30\textwidth]{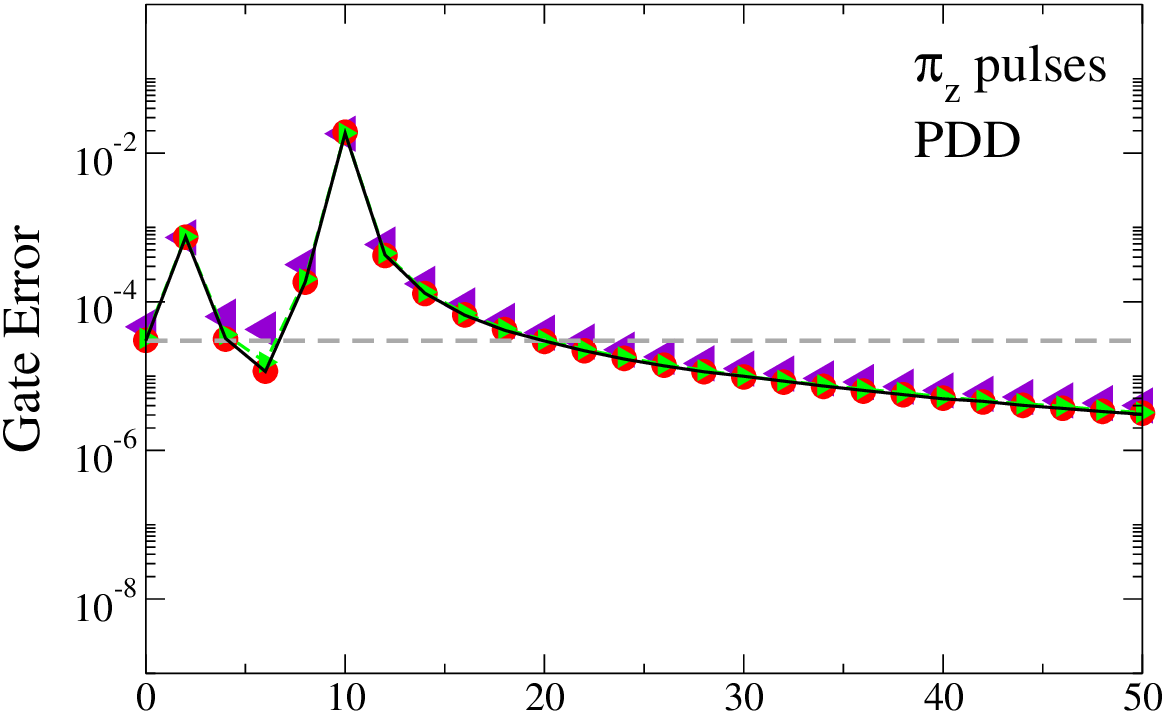} 
\includegraphics[width=0.29\textwidth]{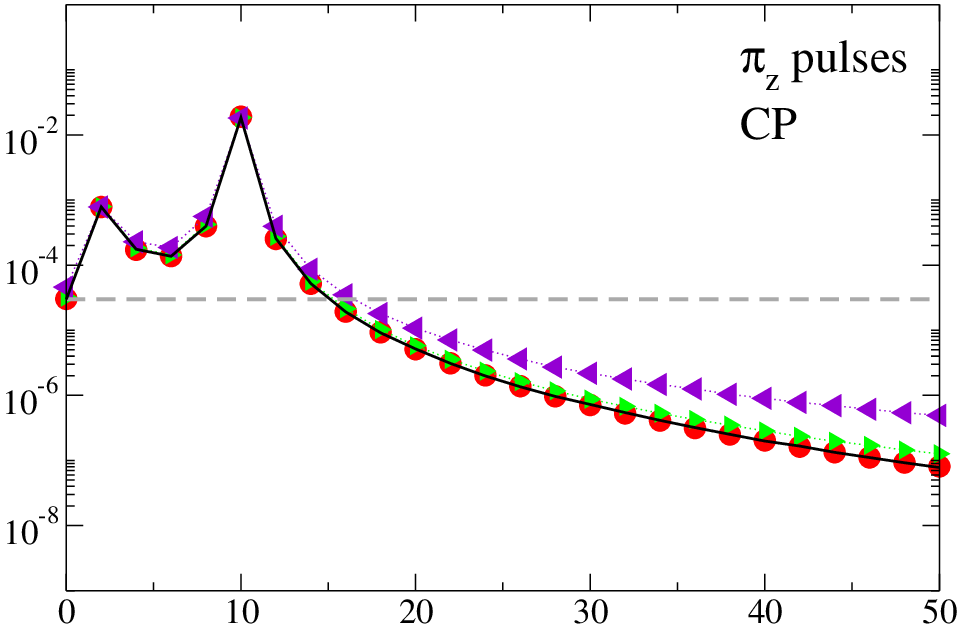}
\includegraphics[width=0.29\textwidth]{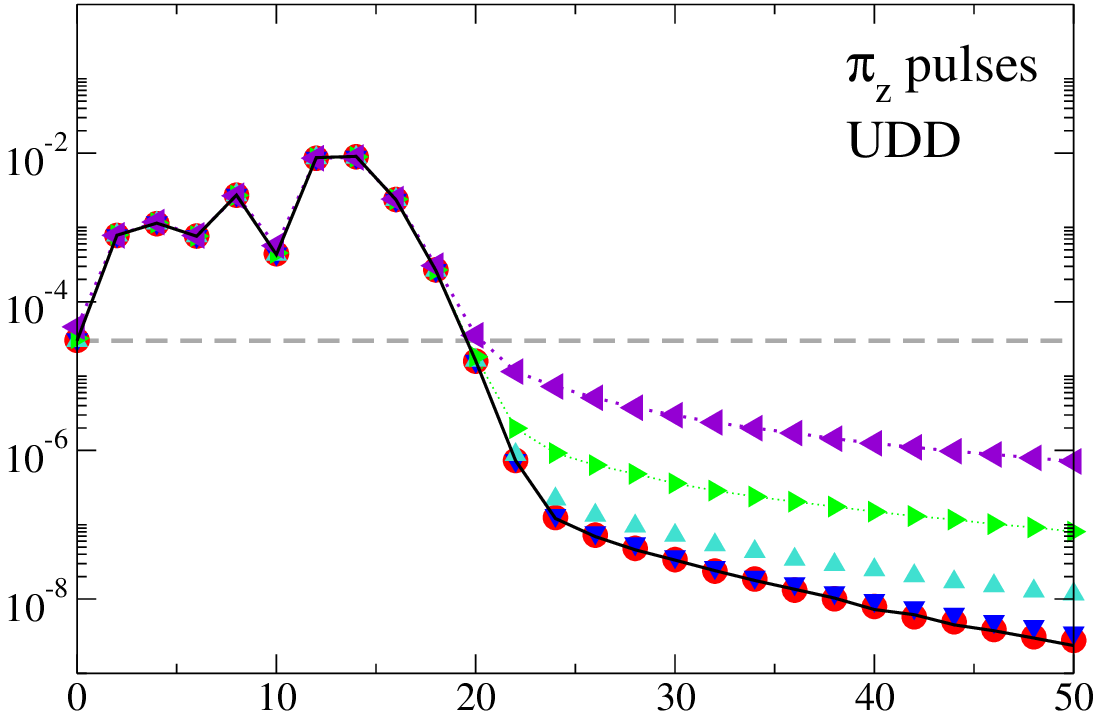} \\
\includegraphics[width=0.30\textwidth]{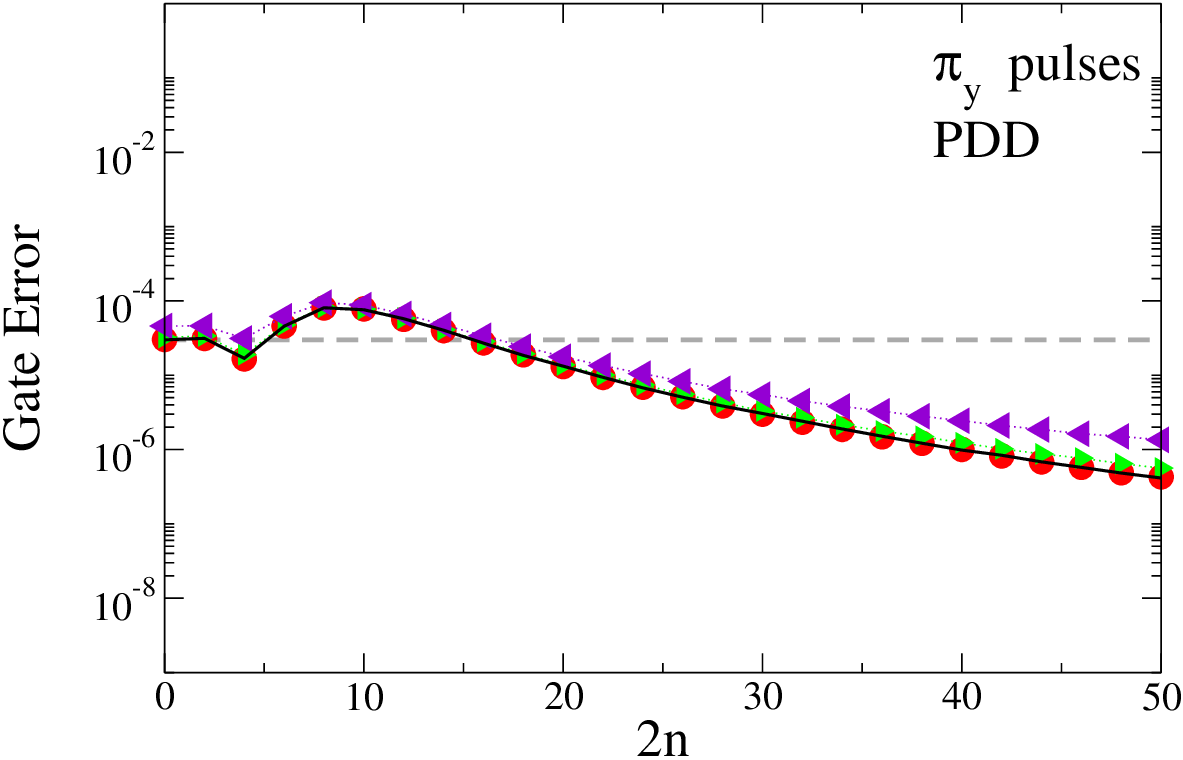} 
\includegraphics[width=0.29\textwidth]{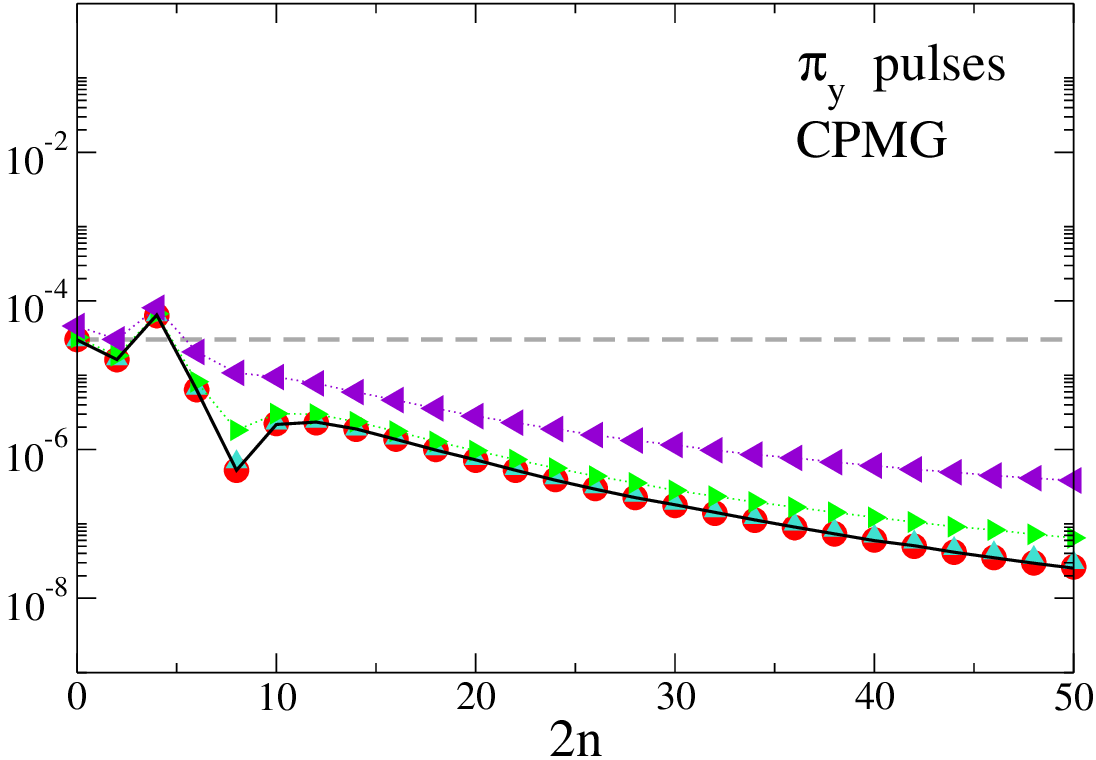}
\includegraphics[width=0.29\textwidth]{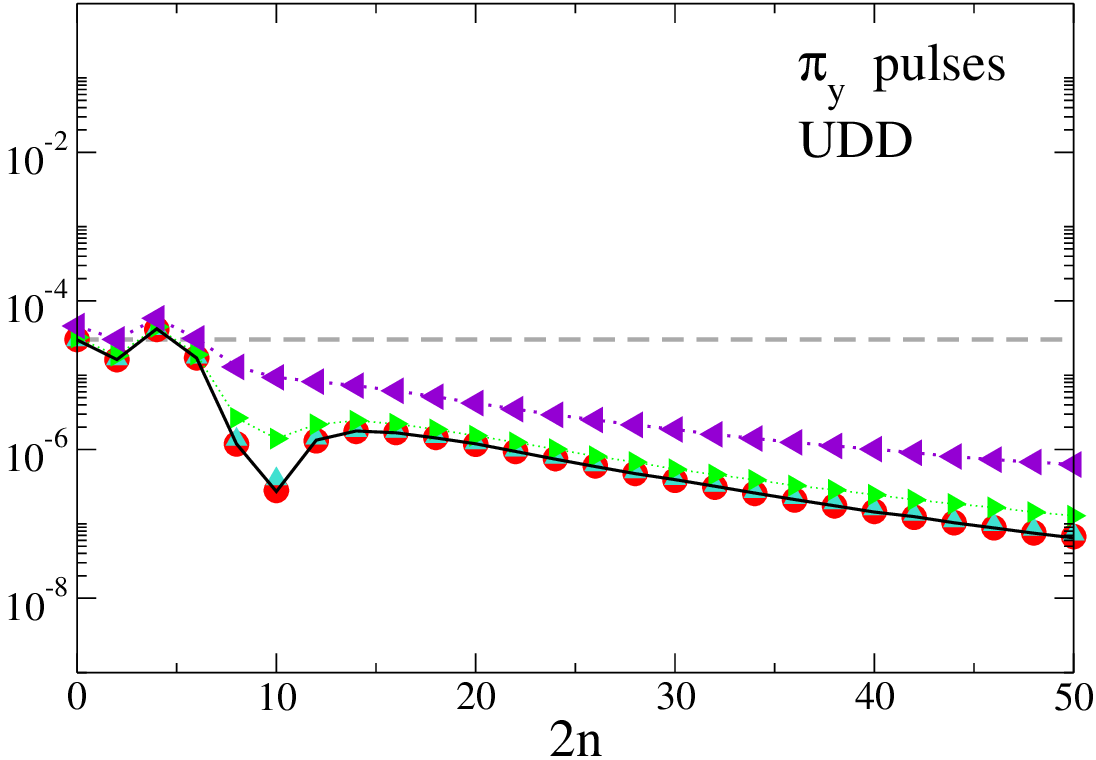}
  \end{center}
  \caption{{\bf{Robustness of the quasi-static approximation with increasing UV cut-off}} (Color online)
Gate error at time $t_e= \pi/(2\omega_c)= 2n \Delta t$ ($n \geq 1$) as a function of 
the number of couple of pulses. The black continuous line is the gate error for quasi-static $1/f$ noises with   
 $\Sigma_1=\Sigma_2=  10^9$~rad/s ($\gamma_m=\gamma_M= 1$s$^{-1}$).
Other symbols correspond to dynamic $1/f$ noise with different UV cut-off:
$\gamma_M=10^6$s$^{-1}$ circles (red),
$\gamma_M=10^{7}$ s$^{-1}$ triangles down (blue),
  $\gamma_M=10^{8}$ s$^{-1}$  triangles up (turquoise),
  $\gamma_M=10^{9}$ s$^{-1}$  triangles right (green), and
  $\gamma_M=10^{10}$ s$^{-1}$  triangles left (violet).
 The gray dashed line marks the gate error in the absence of pulses, i.e. the value for $n=0$,
  for $\gamma_M= 10^{6}$ s$^{-1}$.
Top row: $\pi_z$-pulses for PDD, CP and UDD. Bottom row: $\pi_y$ pulses for PDD, CPMG and UDD. 
 Parameters are $\omega_c= 5 \times 10^9$~rad/s, $\Omega= 10^{11}$rad/s.
 Data obtained from the solution of the SSE. Dotted lines are guides to the eye.
}
  \label{figure2}
\end{figure*}
This expression has to be averaged over the static values $x_i$, describing repeated measurements.
We assume 
independent stochastic processes, $P(x_1,x_2)= P(x_1)P(x_2)$, 
where $P(x_i)$ are Gaussian distributed variables with variance $\Sigma_i$ \cite{falci2005PRL}. 
Considering that $t_e= \pi/2 \omega_c$ and $\omega_c \ll \Omega$ the average error, for $n > n_0$, is found as
\begin{eqnarray}
\langle \varepsilon \rangle_{\small{\rm{PDD}}} &=& {\pi^2 \over 2^7} \frac{\Sigma_1^2+\Sigma_2^2}{\omega_c^2} \frac{1}{n^2} 
\Big [1- {1 \over \sqrt 2} \cos \Big (\sqrt{\Omega^2 + {\omega_c^2 \over 4}}t_e \Big ) \nonumber \\
&-&  {1 \over 2 \sqrt 2} {\omega_c \over \sqrt{\Omega^2 + \omega_c^2/4}} \sin \Big ( \sqrt{\Omega^2 + {\omega_c^2 \over 4}}t_e \Big ) \Big ]
\nonumber \\
&\approx& {\pi^2 \over 2^7} \frac{\Sigma_1^2+\Sigma_2^2}{\omega_c^2} \frac{1}{n^2} \Big [1- {1 \over \sqrt 2} 
\cos \Big ( \frac{\pi \Omega}{2\omega_c} \Big ) \Big ] \, .
\label{av-errorPDD}
\end{eqnarray}
We note that the gate error under PDD scales with $(\Sigma_i/\omega_c)^2$, where $\Sigma_i^2$ is related to the amplitude of $1/f$ noise, 
and that it decreases $\propto n^{-2}$. 

The validity limits of this approximation can be estimated by comparing it with the exact numerical solution of
the SSE (Figs. \ref{figure1}, \ref{figure2}). In Fig. \ref{figure1} 
we report the gate error  at time $t_e= \pi/(2\omega_c)= 2n \Delta t$
as a function of the number of pulses considering $\gamma_M/\gamma_m = 10^6$.
The error for $n=0$ is due to the effect of $1/f$ noise and it is larger the larger is the noise amplitude, i.e. the
variances $\Sigma_i$. The predicted scaling of the gate error as $n^{-2}$ with increasing number of pulses is
confirmed by the solution of the SSE (inset). Moreover, the value $n_0$ resulting from validity condition of the third-order 
truncation of the Magnus expansion,  also gives a threshold value separating a regime of initial non-monotonic behavior 
from a regime where the gate error monotonically decreases.  
We observe in fact that the application of few pulses $n < n_0 \approx 5$, may even increase the gate error 
with respect to the unconditioned evolution. This  is reminiscent of the anti-Zeno effect which occurs under PDD
in a qubit at the optimal point, as reported in \cite{falci2005PRL,falci2005PhysE}. 
The robustness of the quasi-static approximation with increasing high-frequency (soft) 
cut-off $\gamma_M$, for fixed noise variance $\Sigma_i$, is reported in Fig.\ref{figure2} (left panels).
Remarkably, the quasi-static approximation of the gate error under PDD is valid until $\gamma_M =10^{10}$s$^{-1}\simeq \Omega$.

\subsubsection{Carr-Purcell and Uhrig dynamical decoupling}

\begin{figure}[t!]
\begin{center}
{
\includegraphics[width=0.40\textwidth]{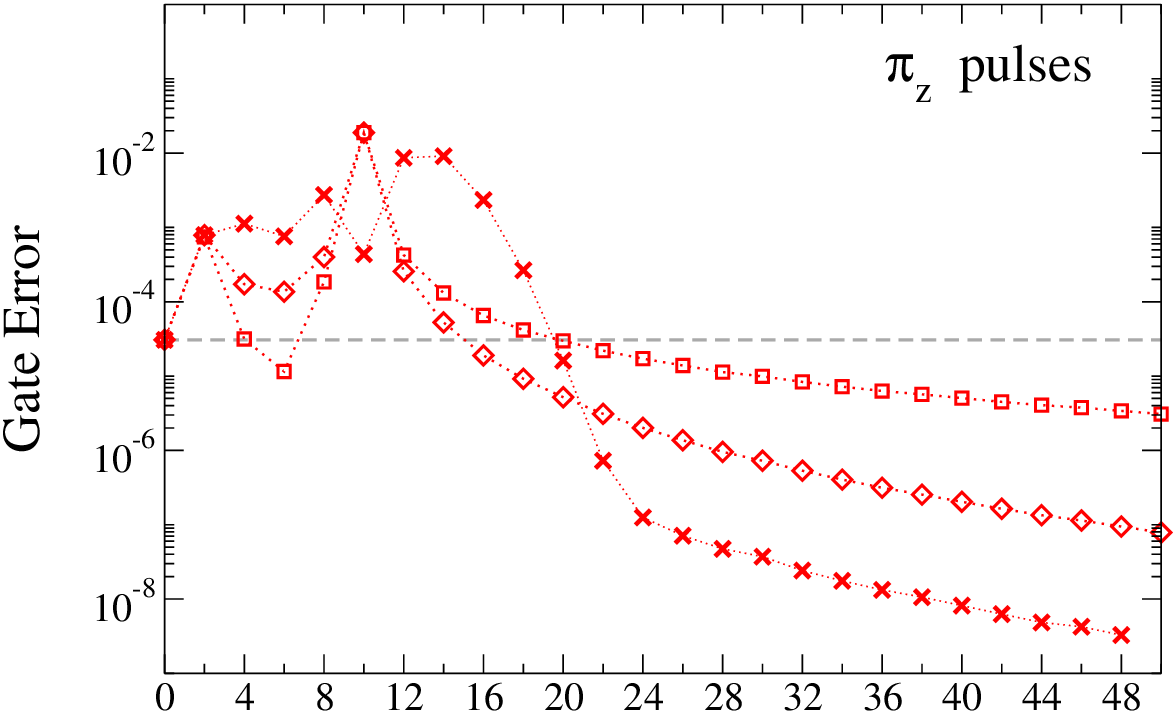} \\
\includegraphics[width=0.40\textwidth]{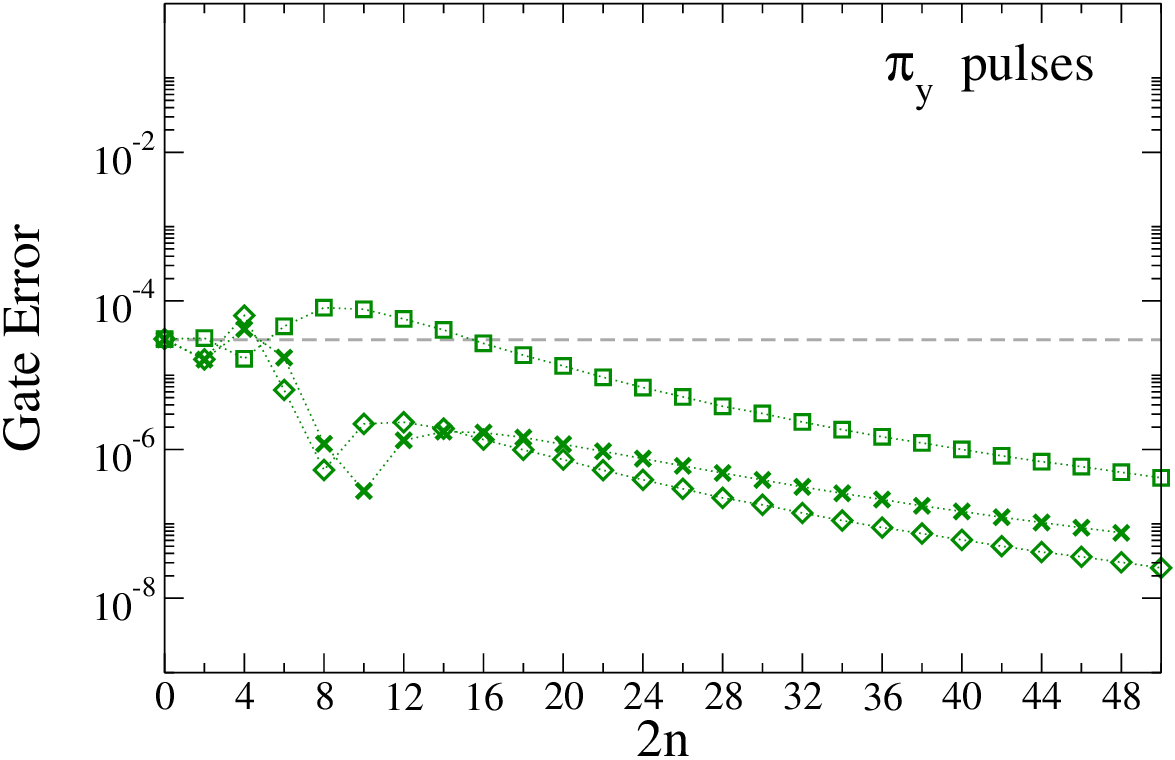}}
\caption{\label{figure3}\footnotesize (Color online)
Gate error at time $t_e= \pi/(2\omega_c)= 2n \Delta t$ ($n \geq 1$) as a function of 
the number of couple of pulses.
Top panel: $\pi_z$-pulses for PDD (squares), CP (diamonds) and UDD (crosses).
Bottom panel: $\pi_y$ pulses for PDD (squares), CPMG (diamonds) and UDD (crosses).
The gray dashed line marks the gate error in the absence of pulses, i.e. the value for $n=0$.
Parameters are $\omega_c= 5 \times 10^9$~rad/s, $\Omega= 10^{11}$rad/s,
$\Sigma_1=\Sigma_2=  10^9$~rad/s, $\gamma_m= 1$s$^{-1}$, $\gamma_M=10^6$s$^{-1}$.
Data obtained from the solution of the SSE. Dotted lines are guides to the eye.  }
\end{center}
\end{figure}

\begin{figure}[t!]
\begin{center}
{
\includegraphics[width=0.40\textwidth]{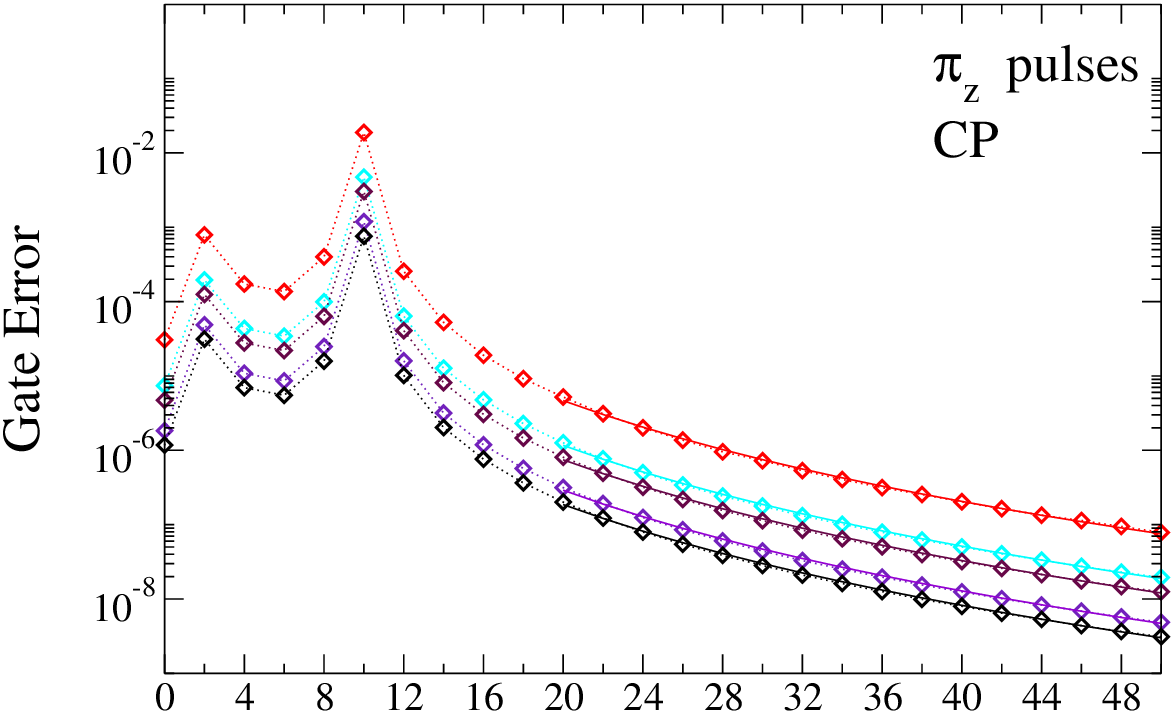} \\ 
\includegraphics[width=0.40\textwidth]{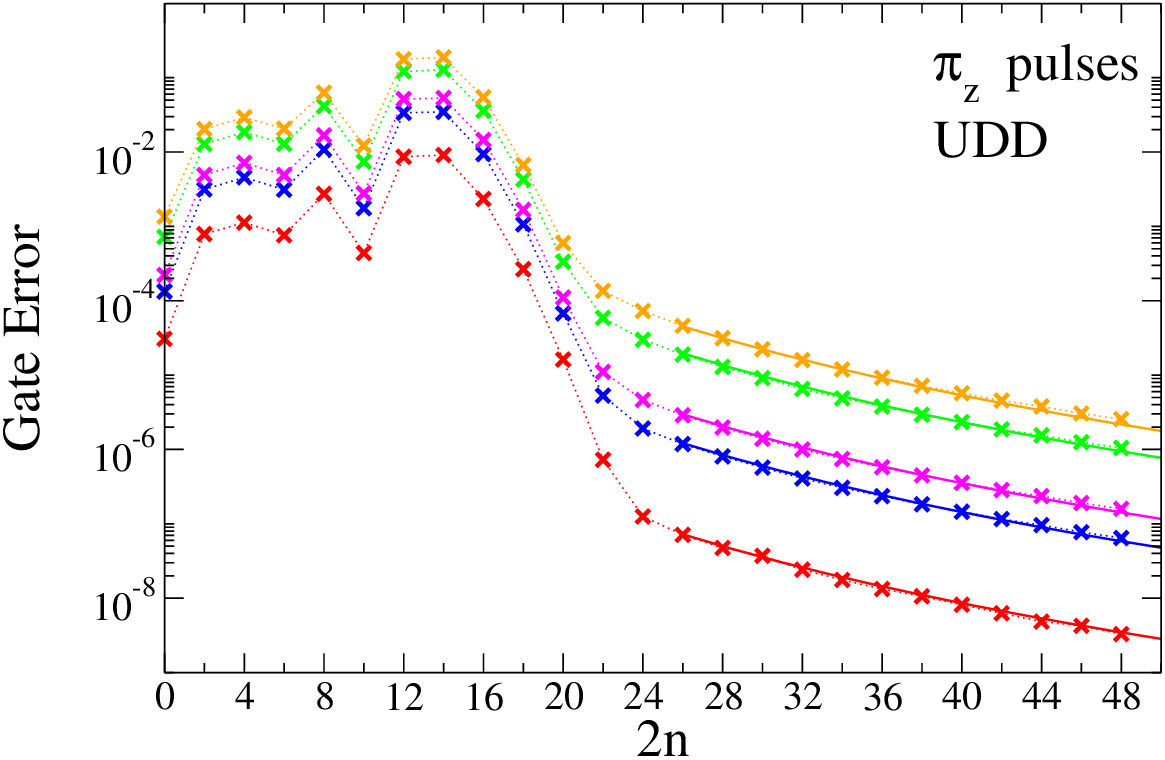}\\ \hspace{0.2cm}
\includegraphics[width=0.39\textwidth]{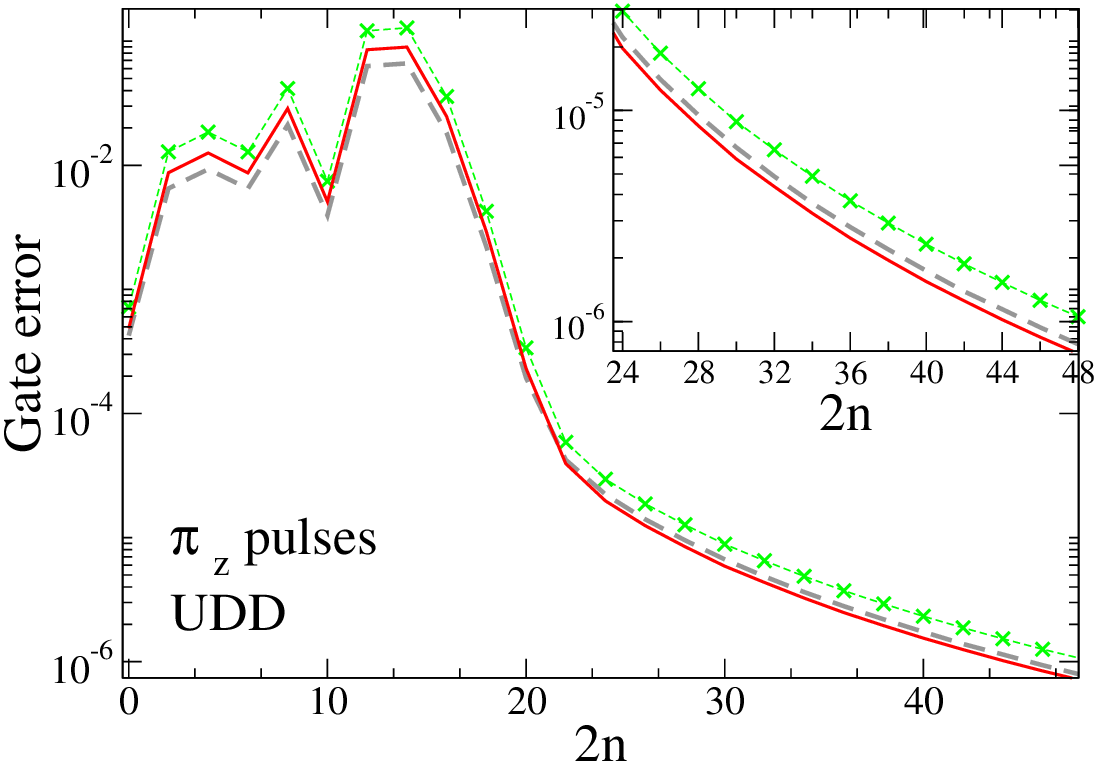}}
\caption{\label{figure4}\footnotesize (Color online)
Scaling of the $n$-dependence of gate error at $t_e$ with the noise amplitudes, $\Sigma_i$, 
for  CP (top panel) and UDD (middle and bottom panels).
Top panel: from top to bottom $\Sigma_1= \Sigma_2=  [1, 0.5, 0.4, 0.25, 0.2] \times 10^9$~rad/s.
The continuous lines ($2n \geq 20$) are fit with $n^{-4.5}$ and scale as $\Sigma_i^2$.
Middle panel: from top to bottom $\Sigma_1= \Sigma_2=  [5, 4, 2.5, 2, 1] \times 10^9$~rad/s.
The continuous lines ($2n \geq 26$) are fit with $n^{-5}$ and scale as $\Sigma_i^4$.
Bottom panel: sensitivity to different noise amplitudes on the two qubits as given by Eq.(\ref{av-errorUDD}) with q=-0.7 for
$\Sigma_1= 4 \times 10^9$, $\Sigma_2=0$ (gray dashed), $\Sigma_1=\Sigma_2= 4 \times 10^9$
(green crosses) and $\Sigma_1=4 \times 10^9$, $\Sigma_2 = 2.4 \times 10^9$ (red continuous). 
Data obtained from the solution of the SSE.
Parameters are $\omega_c= 5 \times 10^9$~rad/s, $\Omega= 10^{11}$rad/s,
and $\gamma_m= 1$s$^{-1}$, $\gamma_M=10^6$s$^{-1}$.}
\end{center}
\end{figure}

\begin{figure}[t!]
\begin{center}
{
\includegraphics[width=0.35\textwidth]{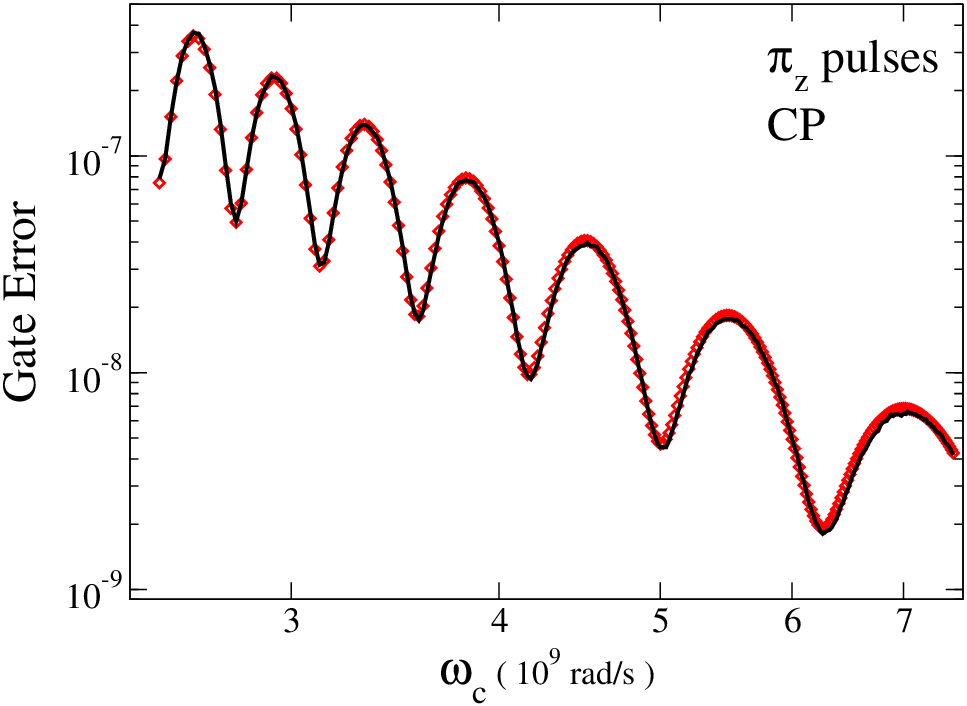} \\ \vspace{0.3 cm}
\includegraphics[width=0.35\textwidth]{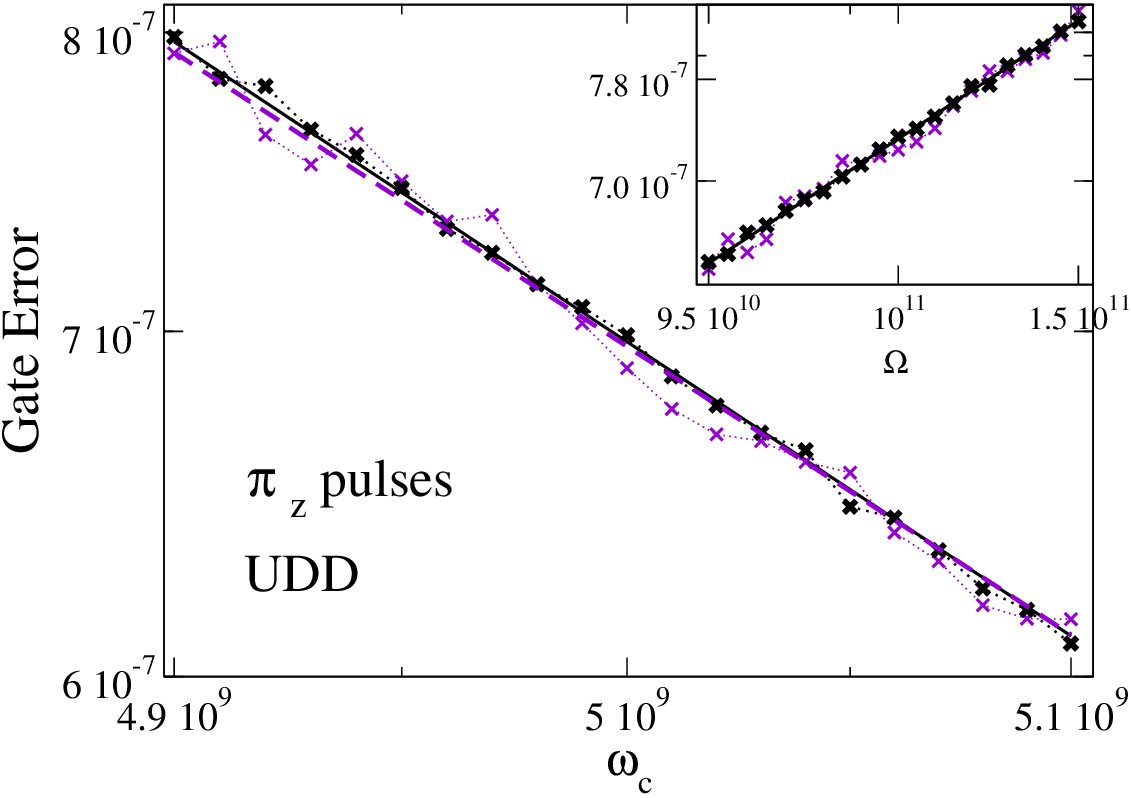}
}
\caption{\label{figure5}\footnotesize (Color online)
Gate error at time $t_e$ for $n=50$ as a function of $\omega_c$ for $\Omega= 10^{11}$rad/s.
Top panel: numerical data for CP (red diamonds) and fit 
with Eq. (\ref{av-errorCP}) (black line), $\Sigma_i=  10^9$~rad/s.  
Bottom panel:
Lines are numerical fits with $\omega_c^{-6.6}$ (continuous black) and $\omega_c^{-6.5}$ (dashed violet), cfr Eq.(\ref{av-errorUDD}). Inset: gate error as a function of $\Omega$ and fit with $\Omega^{2.6}$ (continuous black), here
$\omega_c= 5 \times 10^9$~rad/s. 
Numerical data averaged over different numbers of noise realizations 
$N=10^5$ (violet) and $N=10^6$ (black) for $\Sigma_i=  3 \times 10^9$~rad/s.
In both panels the bandwidth is $\gamma_m= 1$s$^{-1}$, $\gamma_M=10^6$s$^{-1}$. 
} 
\end{center}
\end{figure}

The qualitative features derived for PDD are found also for the Carr-Purcell and Uhrig sequences (Figs. \ref{figure2} and
\ref{figure3}).
Remarkably different scalings with the number of pulses and, in the large-$n$ regime, with the noise amplitude
and with the gate time, i.e. with $\omega_c$, appear, see Figs. \ref{figure4}-\ref{figure5}.

\begin{figure}[t!]
\begin{center}
{
\includegraphics[width=0.40\textwidth]{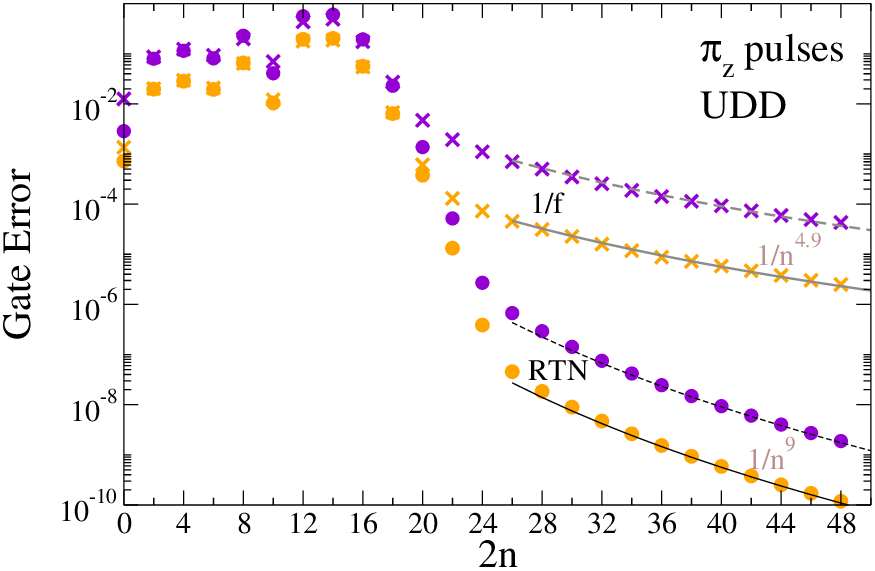}}
\caption{\label{figure6}\footnotesize (Color online)
(Color online) Gate error at time $t_e = \pi/(2\omega_c) = 2n\Delta t$
 ($n \geq 1$) as a function of the number of couple of pulses. The points
 at $n = 0$ represents the gate error in the absence of DD.
 Crosses correspond to $1/f$ noise with $\gamma_m=\gamma_M=1$ s$^{-1}$,
 upper symbols (violet) refer to
 $\Sigma_i = 10^{10} \equiv \Sigma^A$ rad/s, whereas lower symbols (orange)
 refer to  $ \Sigma_i = 5\cdot 10^9 \equiv \Sigma^B$ rad/s.
 Circles correspond to RTN noise $\{-v/2,v/2\}$ with 
 $\gamma_0=1$ s$^{-1}$, upper symbols (violet) refer to
 $v_A = 2\cdot10^{10}$ rad/s, whereas lower symbols (orange)
 refer to  $v_B = 10^{10}$ rad/s. These values are chosen to ensure that the RTN and $1/f$ noise have equal
 first and second order momenta. 
 The (black) continuous thin curve is a numerical fit $\propto n^{-9}$, multiplied by $(v_B/v_A)^4 = 2^4$ 
 one obtains the (black) dashed thin curve.
 The (gray) continuous thick curve is a numerical fit $\propto n^{- 5}$,  multiplied by $(\Sigma_B/\Sigma_A)^4 = 2^4$ 
 one gets the (gray) dashed thick curve. 
 Parameters
 are $\omega_c = 5 \cdot 10^9$ rad/s, $\Omega = 10^{11} $rad/s.
}
\end{center}
\end{figure}

For small number of pulses, the error is non-monotonic in $n$ until a sequence-specific threshold value $n_0$ scaling with 
$\propto \omega_c^{-1}$. For the CP sequence the crossover takes place at $n_0$ found for the PDD, for UDD the crossover to a fast monotonic 
decay  takes place at a larger value of $n$,  Fig. \ref{figure3} (top panel). 
For intermediate number of pulses CP decoupling slightly reduces the error with respect to PDD and UDD. 
For large pulse numbers instead, UDD outperforms PDD and CP.
A numerical fit of the gate error in the asymptotic regime gives respectively for CP and UDD
\begin{eqnarray}
\label{av-errorCP}
&& \!\!\!\! \!\! \!\!\!\! \!\!  \langle \varepsilon \rangle_{\rm{CP}}  \approx \frac{\Sigma_1^2+\Sigma_2^2}{\omega_c^2} 
\Big (\frac{\Omega}{\omega_c} \Big )^2 \frac{1}{n^{\alpha_{\rm{CP}}}} 
\Big [1- {1 \over \sqrt 2} \cos \Big ( \frac{\pi \Omega}{2\omega_c} \Big ) \Big] \, , \\
\label{av-errorUDD}
&&  \!\!\!\! \!\! \!\!\!\! \!\! \langle \varepsilon \rangle_{\rm{UDD}}  \approx \frac{\Sigma_1^4+\Sigma_2^4 + q \Sigma_1^2 \Sigma_2^2}{\omega_c^{4}} 
\Big ( \frac{\Omega}{\omega_c} \Big )^{2.6} \frac{1}{n^{\alpha_{\rm{UDD}}}} \, ,
\end{eqnarray}
with $\alpha_{\small{\rm{CP}}} \approx 4$ and $\alpha_{\small{\rm{UDD}}} \approx 5$.
The dependences on $\Sigma_i$ and $n$ are shown in Fig. \ref{figure4}, the $\omega_c$ dependence 
is reported in Fig. \ref{figure5}.  
The gate error under CP has the same periodicity with $\omega_c$ found for the PDD, Eq. (\ref{av-errorPDD}), the more rapid decrease 
with increasing qubit's coupling under CP is due to the extra factor $(\Omega/\omega_c)^2$ in Eq. (\ref{av-errorCP}). 
The gate error under UDD instead is not periodic with $\omega_c$,  Fig. \ref{figure5}.
As a difference with PDD and CP, the asymptotic error under UDD scales with $\Sigma_i^4$. 
Moreover, due to the cross term $ q \Sigma_1^2 \Sigma_2^2$
with $q<0$, the gate efficiency may be larger when both $\Sigma_i \neq 0$ than when only one qubit is noisy 
(gray dashed line and green crosses in Fig.(\ref{figure4}) bottom panel).
The dependence on $\Sigma_i^4$ 
in  Eq. (\ref{av-errorUDD}) is an indication 
that the leading order contribution to the error is due the fourth order statistics 
of the stochastic process and suggests that the second order contribution
to the error is suppressed by the Urhig sequence. 
For a single qubit at pure dephasing (longitudinal noise) this is an established result. In fact, the UDD sequence has been designed to eliminate the effect of longitudinal Gaussian noise to lowest order \cite{uhrig2007PRL}.  
We remark that we are not considering the generalized Urigh sequence introduced in \cite{Yang2008}
to suppress relaxation (transverse noise).
The contribution of fourth-order statistics is highlighted comparing 
the gate error under UDD for qubits affected by $1/f$ noise and
by a symmetric random telegraph process $\xi(t)$ switching between $\pm v/2$ with rate $\gamma_0$ \cite{DArrigoPhysScr2015}. 
The numerical analysis is reported in Fig. \ref{figure6}.
We consider the limiting case of static noise with respect to the gate operation, i.e. we fix the switching rates  
$\gamma_0=\gamma_m=\gamma_M= 1$s$^{-1}$, and take the two processes with equal
second order statistics, i.e. $\Sigma^2= \langle \xi^2 \rangle = (v/2)^2$ (the average vanishes for both
processes).  
For small pulse numbers the dominant contribution to the gate error comes from the second moment, in fact the gate error is nearly the same for RTN and $1/f$ noise. Few pulses are not sufficient to eliminate the Gaussian
contribution. For larger pulse numbers instead, when UDD becomes effective, the procedure is sensitive to the noise statistics. The gate error differs in the two cases and 
scales  $\propto \Sigma^4$ in the presence of $1/f$ noise and  $\propto v^4$ in the presence of RTN. 
This suggests a dependence of the gate error on the fourth noise cumulant, which differs for the two processes, being
$\langle \langle x_{i1} x_{i2}x_{i3}x_{i4} \rangle \rangle=0$ for Gaussian static $1/f$ noise and
 $\langle \langle \xi_{i1} \xi_{i2} \xi_{i3} \xi_{i4} \rangle \rangle= - v^4/8$ for static RTN.
Note also the different dependence on the number of applied pulses, with the very fast decay $\propto n^{-9}$ in the presence of RTN 
\cite{DArrigoPhysScr2015}.

The robustness of the quasi-static approximation with increasing UV cut-off is illustrated in  Fig. \ref{figure2}.  
Similarly to PDD, the gate error under CP is weakly dependent on $\gamma_M$ (the quasi-static approximation is valid 
until $\gamma_M \leq 10^9$s$^{-1}$). 
On the other side, the gate performance under UDD is the very sensitive to the high-frequency (soft) cut-off of the noise,  analogously to other analysis \cite{cywinsky2008PRB,LoFrancoPRB2014}.

\subsection{Dynamical Decoupling with $\pi_y$ pulses}

We already noticed that for quasi-static noise simultaneous $\pi_y$-pulses also cancel the individual qubit Hamiltonian, 
$-\Omega \sigma_z/2$, leading to an effective evolution under $\mathcal{\tilde H} = (\omega_c/2) \sigma_{1x} \otimes \sigma_{2x}$.
As long as the system, prepared in $| +- \rangle$, remains in the SWAP subspace, $\pi_z$ and $\pi_y$ sequences are equivalent. On the other side, transverse noise on each qubit mixes the SWAP and the orthogonal subspace $\{|0\rangle, |3\rangle\}$ (see Table I) \cite{Paladino2011},
making pulses along the two axis inequivalent.
An indication of this fact for PDD comes from the time evolution operator Eq.(\ref{eq:propagator}) evaluated
making the Magnus expansion which, up to the third order, can be expressed in terms of
\begin{eqnarray}
\mathcal H_{eff}(\Delta t) &=&  \tilde{\mathcal H}+ 
\frac{\omega_c \Delta t}{2}  \frac{\Omega}{8}[ \sigma_{1y} \otimes  \sigma_{2x} + \sigma_{1x} \otimes  \sigma_{2y} \nonumber \\ 
&+& \frac{\Omega \Delta t}{3} (\sigma_{1x} \otimes  \sigma_{2x} - \sigma_{1y} \otimes  \sigma_{2y} ) ] \nonumber \\
&+& \frac{\omega_c \Delta t^2}{6}  \frac{\Omega}{8}[ x_1 \sigma_{1z} \otimes  \sigma_{2x} + x_2 \sigma_{1x} \otimes  \sigma_{2z}]
\label{Magnus-3y}
\end{eqnarray}
where we used the condition $t_e= 2 n \Delta t$. The lowest order corrections to $\tilde{\mathcal H}$ mix states
inside the SWAP subspace. The terms containing noise instead connect the SWAP and the orthogonal subspace. 
Moreover, noise terms are of order $\Delta t^2$, as a difference with $\pi_z$ pulses where noise terms
are linear in $\Delta t$. 
\begin{figure}[t!]
\begin{center}
{
\includegraphics[width=0.40\textwidth]{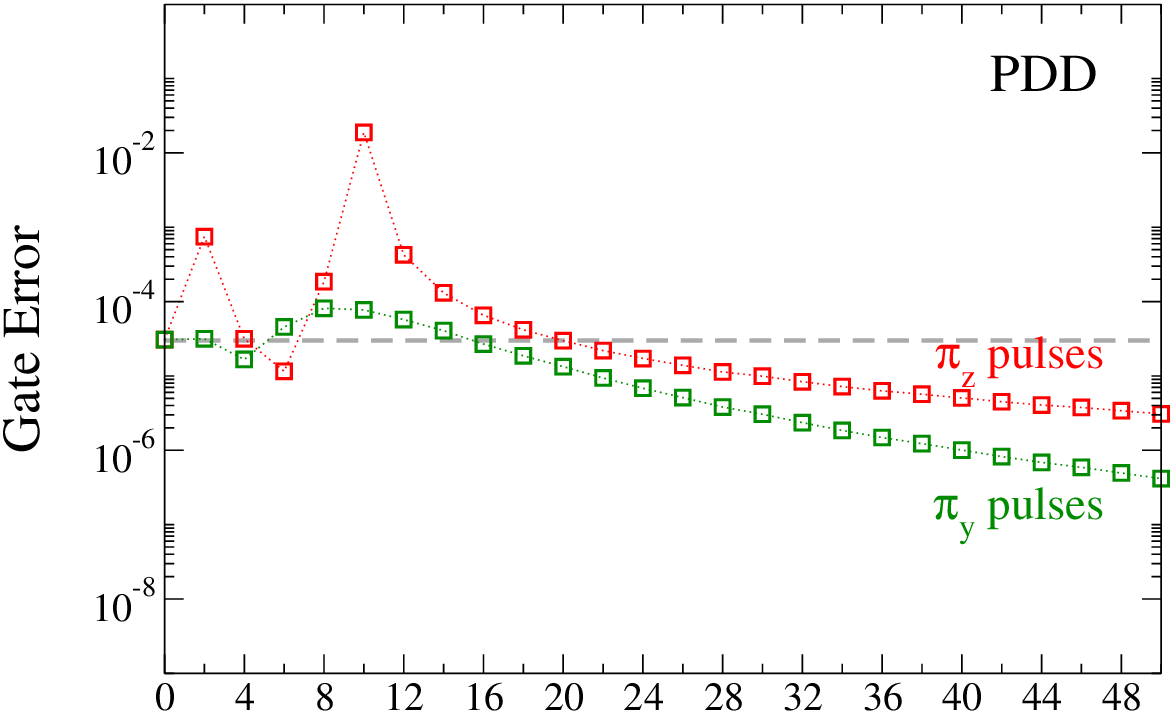} \\
\includegraphics[width=0.40\textwidth]{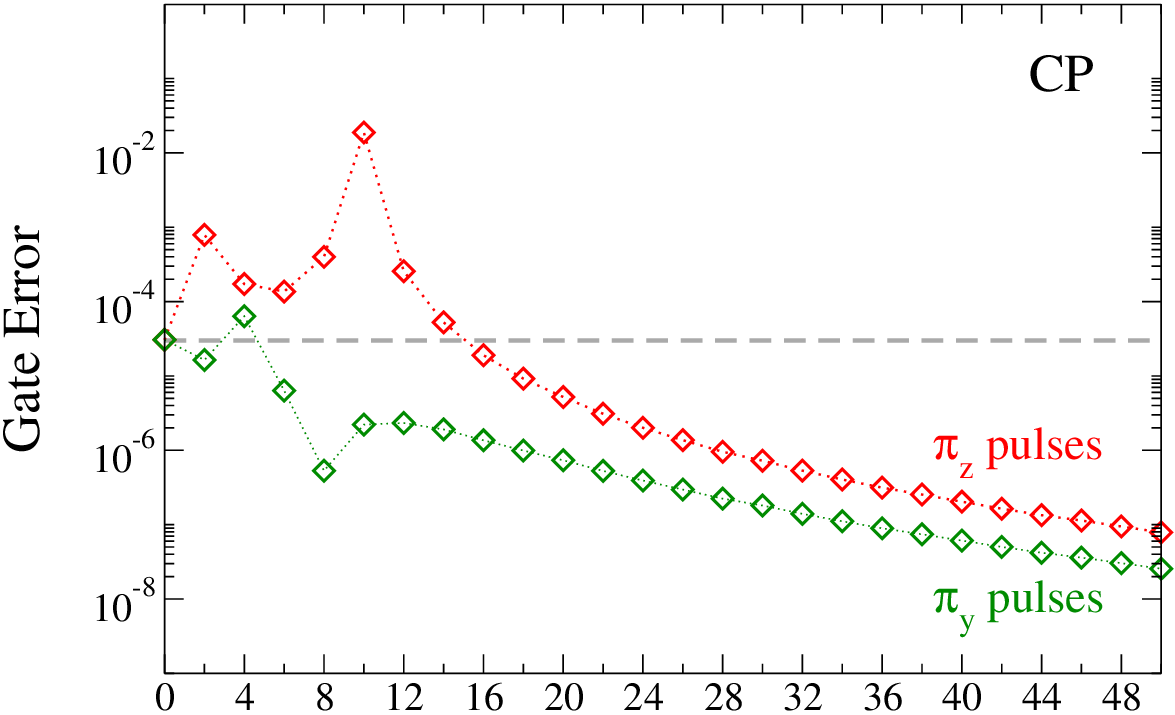}\\
\includegraphics[width=0.40\textwidth]{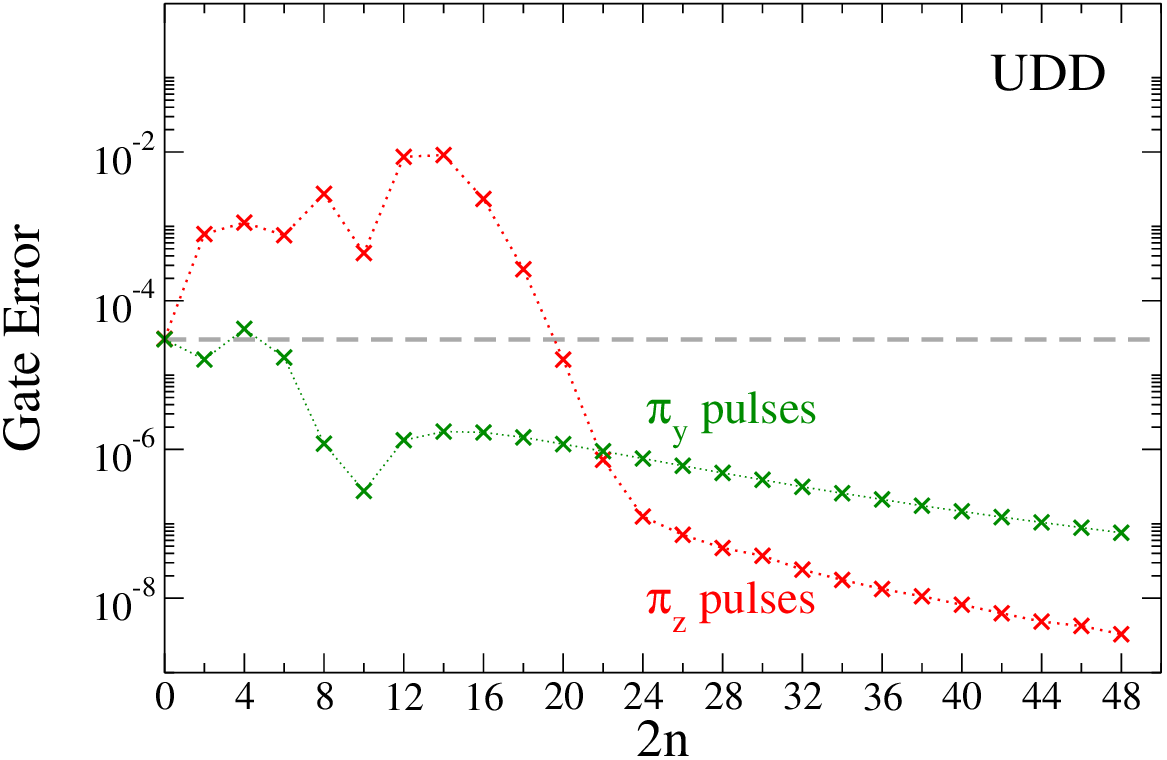}   
}
\caption{\label{figure7}\footnotesize (Color online)
Gate error at time $t_e= \pi/(2\omega_c)= 2n \Delta t$ ($n \geq 1$) as a function of 
the number of couple of $\pi_z$ and $\pi_y$ pulses for PDD (top panel), CP and CPMG (central panel) and 
UDD (bottom panel).
The gray dashed line marks the gate error in the absence of pulses, i.e. the value for $n=0$.
Parameters are $\omega_c= 5 \times 10^9$~rad/s, $\Omega= 10^{11}$rad/s,
$\Sigma_1=\Sigma_2=  10^9$~rad/s and $\gamma_m= 1$s$^{-1}$, $\gamma_M=10^6$s$^{-1}$. Data obtained from the solution of the SSE. }
\end{center}
\end{figure}

\begin{figure}[t!]
\begin{center}
{
\includegraphics[width=0.35\textwidth]{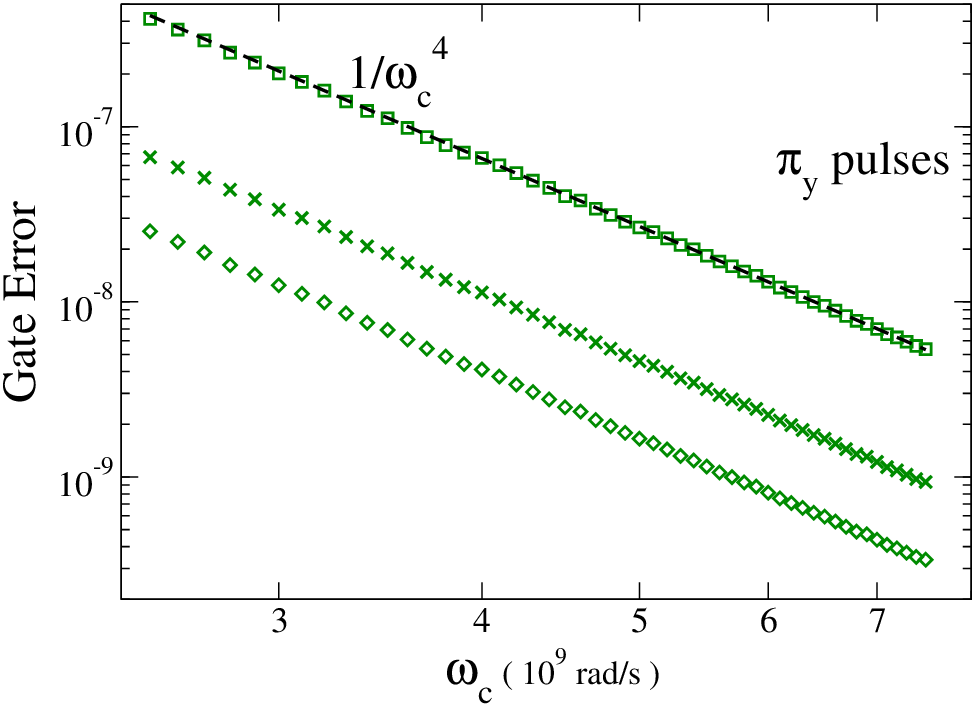} \\ \vspace{0.3 cm}
\includegraphics[width=0.35\textwidth]{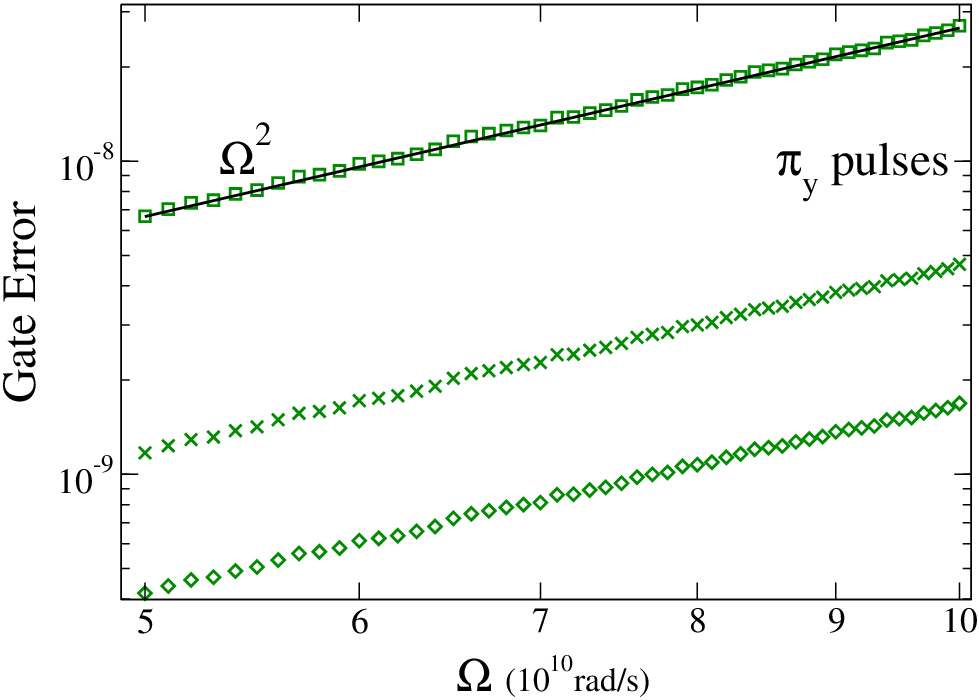} \\ 

}
\caption{\label{figure8}\footnotesize (Color online)
Gate error at time $t_e= \pi/(2\omega_c)= 2n \Delta t$ for large pulse numbers $n=50$ 
as a function of $\omega_c$ for $\Omega= 10^{11}$rad/s (top panel) and
as a function of $\Omega$ for $\omega_c= 5 \times 10^9$~rad/s (bottom panel),
under $\pi_y$ sequences: PDD (squares), CPMG (diamonds), UDD (crosses). The black dashed 
lines are $\propto \omega_c^{-4}$ (top panel) and $\propto \Omega^{2}$ (bottom panel). 
Parameters are $\Sigma_i=  10^9$~rad/s, $\gamma_m= 1$s$^{-1}$, $\gamma_M=10^6$s$^{-1}$. 
} 
\end{center}
\end{figure}

In Fig. \ref{figure7} we compare the effect of $\pi_z$ (red) and $\pi_y$ (green) pulses applied at the same times 
$t_i$, the three panels correspond to PDD, CPMG and UDD.
For small number of pulses we still observe a non-monotonic behavior of the gate error. The threshold value of $n$ is approximately $n_{th}\propto\omega_c^{-1}$ and it does not depend on $\Sigma_i$. 
$\pi_y$-PDD pulses reduce the anti-Zeno behavior observed for $\pi_z$-PDD. 
The CPMG and $\pi_y$-UDD sequences instead do not show anti-Zeno behavior, rather we note that for pulse numbers close to the value of $n_{thr}$ found for $\pi_z$ pulses, the error is minimum instead of maximum.
For all sequences, the error in the asymptotic regime, $n \geq 50$, is fitted by  
\begin{equation}
\langle \epsilon \rangle_{\pi_y} \approx \frac{\Sigma_1^2 + \Sigma_2^2}{\omega_c^2} \Big ( \frac{\Omega}{\omega_c} \Big )^2 \frac{1}{n^{\alpha_y}} \, ,
\label{scalings_piy}
\end{equation}
where $\alpha_y \approx 4$, see Fig. 
 \ref{figure8}.

For the $\pi_y$-PDD and CPMG sequences, operating with $\pi_y$ pulses reduces the gate error, with respect to
$\pi_z$-PDD and CP, for any number of applied pulses. For the UDD instead, an advantage is achieved for 
small/intermediate pulse numbers. For large $n$ instead, operating with $\pi_z$ pulses is more advantageous.
This is due to the fact that the error under $\pi_z$-UDD depends on the fourth order statistics, whereas under $\pi_y$-UDD the error scales with $\Sigma_i^2$, signaling a dependence on the second cumulant.

The quasi-static approximation for  $\pi_y$-PDD and CPMG does not reproduce quantitatively the errors when
high-frequency components are included in the spectrum, Fig. \ref{figure2} bottom panels. The approximation 
is tenable until $\gamma_M \leq 10^7$s$^{-1}$, as a difference with the static approximation for sequences realized with 
$\pi_z$ pulses.
On the contrary, $\pi_y$-UDD has a reduced sensitivity on $\gamma_M$, compared with $\pi_z$-UDD. 

\begin{table}[t!]
{\begin{tabular}{cccc}
\toprule
Pulses & PDD & CP/CPMG & UDD\\
\colrule
$\pi_z$ & 
$(\Sigma/\omega_c)^2 n^{-2}$ 
&  
$(\Sigma \Omega/\omega_c^2)^2 \,  n^{-4}$  & 
$(\Sigma^4 \Omega^{2.6}/\omega_c^{6.6}) \, n^{-5}$ \\
$\pi_y$ & 
 $(\Sigma \Omega/\omega_c^2)^2 \,  n^{-4}$  & 
$(\Sigma \Omega/\omega_c^2)^2 \,  n^{-4}$
&  
$(\Sigma \Omega/\omega_c^2)^2 \,  n^{-4}$ \\
\botrule
\end{tabular}}
\caption{\label{tab:scalings} \footnotesize
Scaling of the gate error with the number of pulses (monotonic region, $n> n_{thr}$), 
with the noise amplitude and coupling energy $\omega_c$ for the indicated sequences 
and $\pi_z$ or $\pi_y$ pulses.
}
\end{table} 

\section{Discussion}
\begin{figure*}[t!]
  \begin{center}
\includegraphics[width=0.32\textwidth]{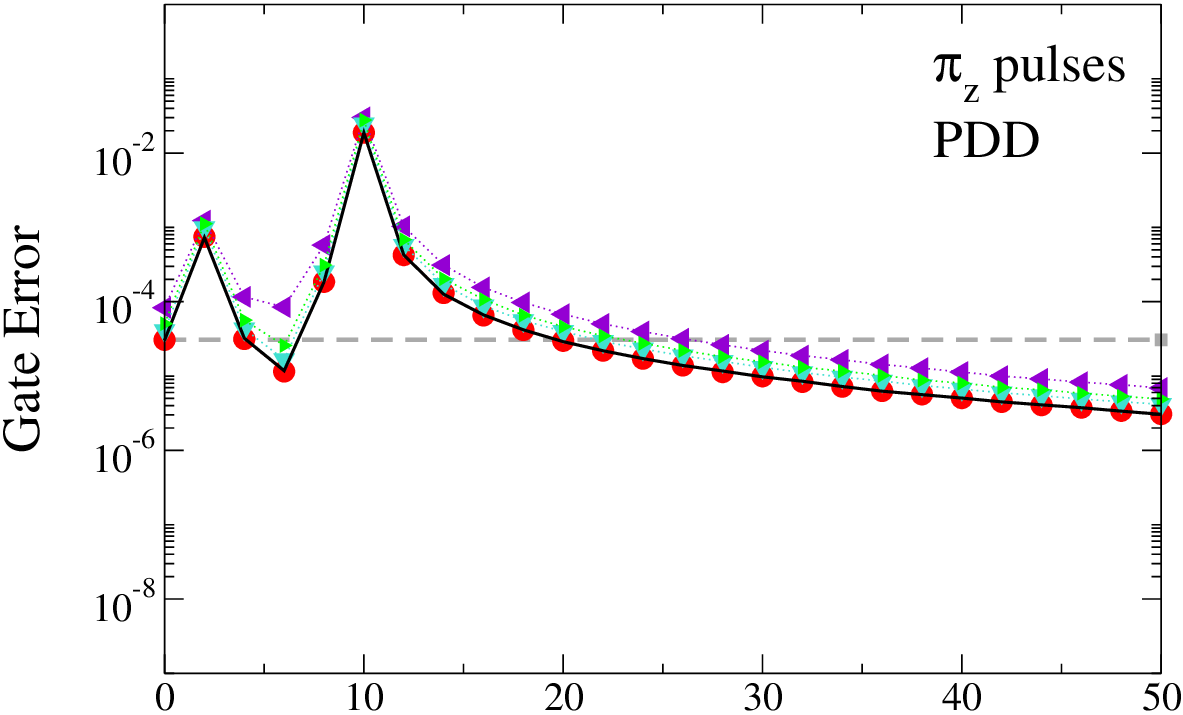} 
\includegraphics[width=0.30\textwidth]{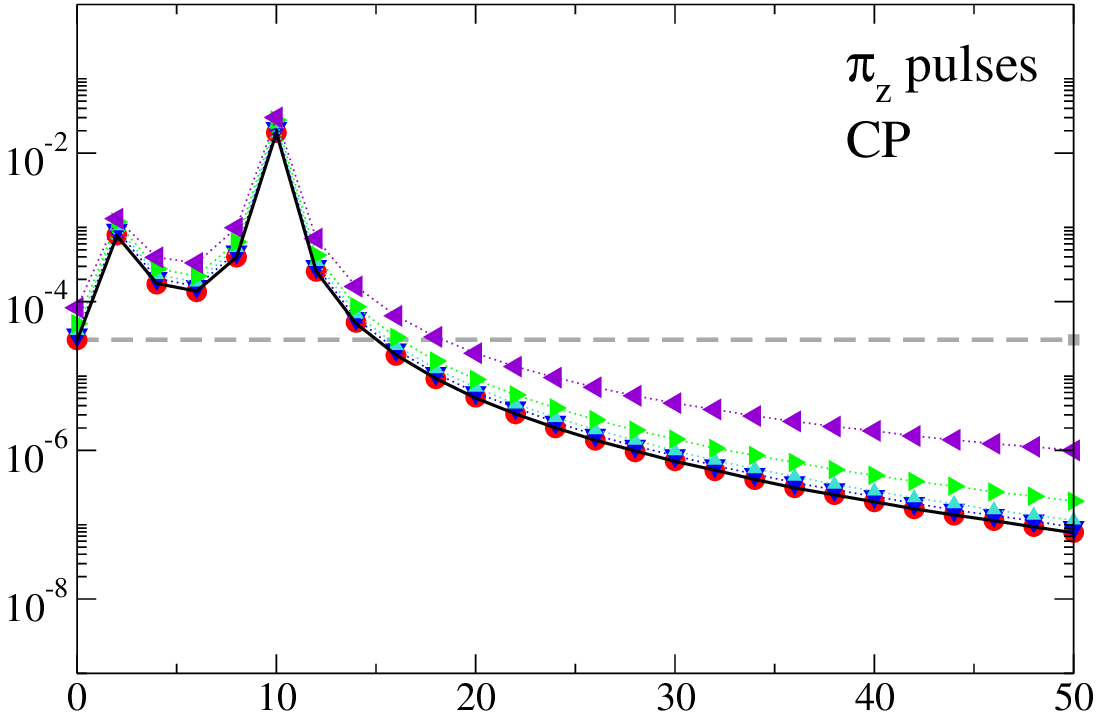}
\includegraphics[width=0.30\textwidth]{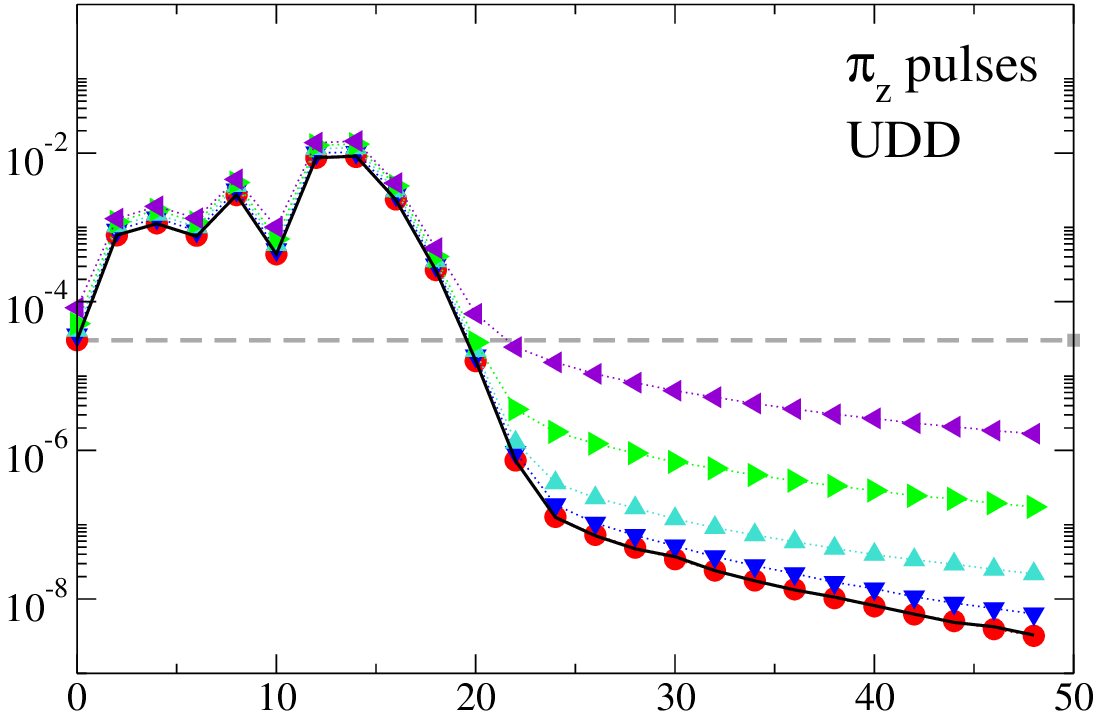} \\
\includegraphics[width=0.32\textwidth]{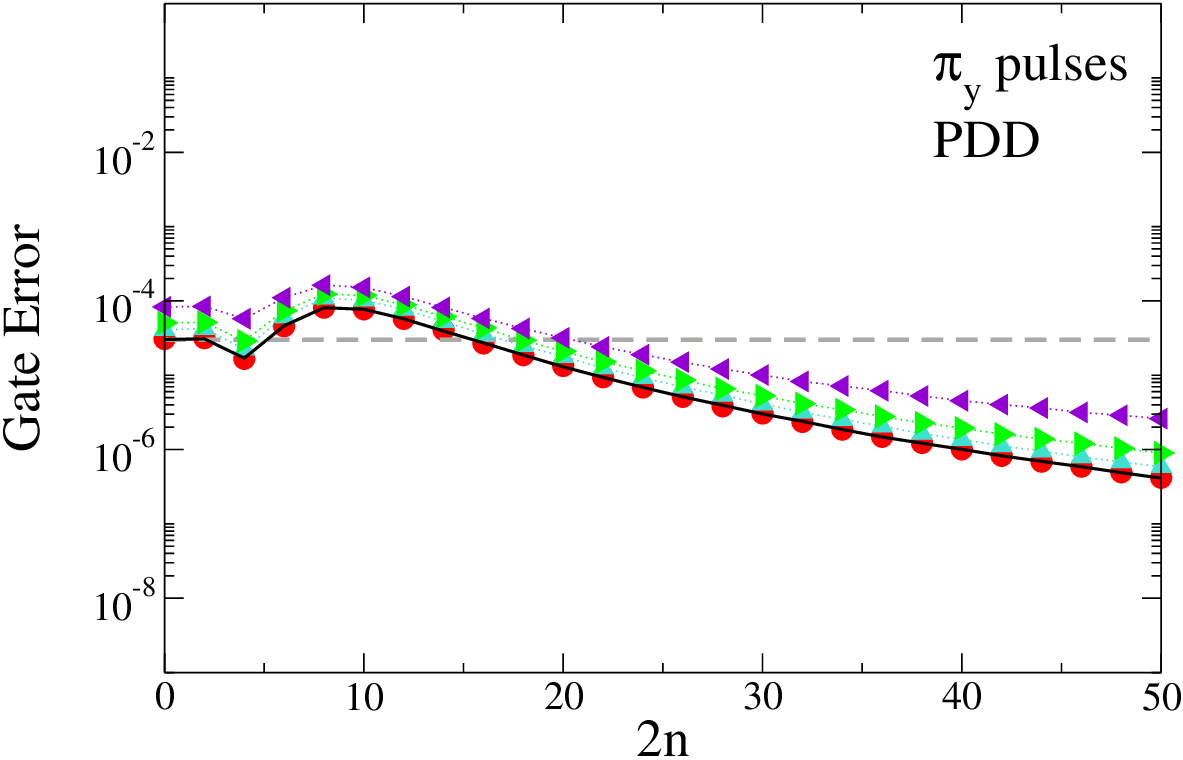} 
\includegraphics[width=0.30\textwidth]{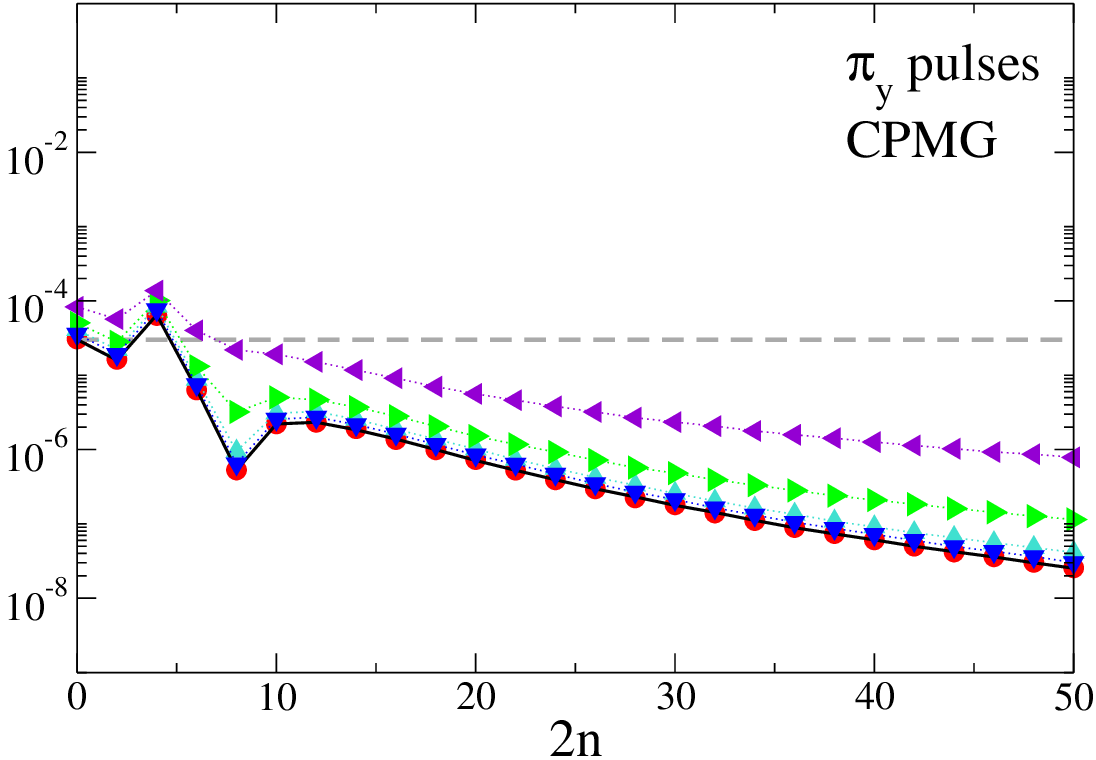}
\includegraphics[width=0.30\textwidth]{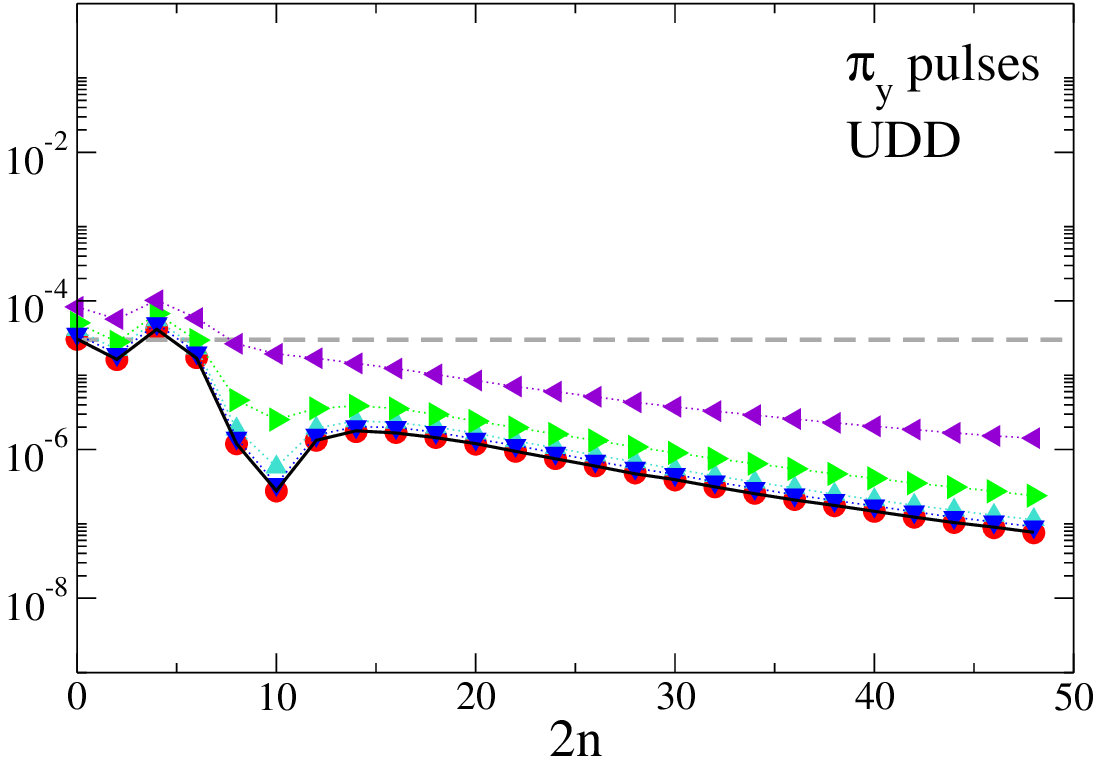}
  \end{center}
  \caption{{ \bf{Effectiveness of decoupling for ''dynamic" $1/f$ noise:}} (Color online)
Gate error at time $t_e= \pi/(2\omega_c)= 2n \Delta t$ ($n \geq 1$) as a function of 
  the number of couple of pulses.
  The filled (red) circles corresponds to $1/f$ noises with   
  $\Sigma_1=\Sigma_2=  10^9$~rad/s, $\gamma_m= 1$s$^{-1}$, $\gamma_M=10^6$s$^{-1}$.
  Other symbols correspond to $1/f$ noises, with the same $\gamma_m$ but increasing
  upper cut-offs (simulation with different number of random telegraph processes to keep
$\Sigma^2/\ln(\gamma_M/\gamma_m)$  constant \cite{PaladinoRMP}): $\gamma_M=10^{7}$ s$^{-1}$  down triangles (blue),
  $\gamma_M=10^{8}$ s$^{-1}$ up triangles  (turquoise),
  $\gamma_M=10^{9}$ s$^{-1}$  right triangles (green), and
  $\gamma_M=10^{10}$ s$^{-1}$  left triangles (violet).
Top row: $\pi_z$-pulses for PDD (left), CP (middle) and UDD (right).
  Bottom row: $\pi_y$ pulses for PDD (left), CPMG (middle) and UDD (right).
 The gray dashed line marks the gate error in the absence of pulses, i.e. the value for $n=0$,
  for $\gamma_M= 10^{6}$ s$^{-1}$.
  The black continuous line corresponds to a \textit{static} $1/f$ noises with   
  $\Sigma_1=\Sigma_2=  10^9$~rad/s ($\gamma_m=\gamma_M= 1$s$^{-1}$). 
Parameters are $\omega_c= 5 \times 10^9$~rad/s, $\Omega= 10^{11}$rad/s.
  Data obtained from the solution of the SSE. Dotted lines are guides to the eye.}
  \label{fig:cutoff-analysis}
\end{figure*}

In this Section we discuss the stability of the various decoupling sequences when decades of high-frequency 
$1/f$ noise are progressively added to the spectrum keeping unchanged the low-frequency part~\cite{paladino2002PRL}. 
This analysis, reported in Fig. \ref{fig:cutoff-analysis}, is done by increasing $\gamma_M$ with fixed noise amplitude $\mathcal A$,
i.e. adjusting the noise variance. 
The expected qualitative feature is that for all sequences part of the efficiency is lost when noise at
higher frequency is present. In the asymptotic limit of large pulse numbers, the $\pi_z$-UDD scheme 
is the most sensitive. Still this procedure guarantees about three-orders of magnitude decrease of the average
error with respect to the unconditioned evolution until $\gamma_M \approx 10^8$s$^{-1}$. 
This is a remarkable result, considering that at present $1/f$ noise has not been detected for frequencies higher 
than $\gamma_M \approx 20$MHz \cite{bylander2011NatPhys,yan2012PRB,PaladinoRMP}.
For smaller pulse numbers, all sequences are less sensitive to high-frequency noise components. In order to avoid anti-Zeno behavior, $\pi_y$ pulses turn out to be more convenient.
Moreover, they also decouple sources of additional noise, longitudinal at the optimal point, as 
$1/f$ critical current fluctuations in flux qubits~\cite{bylander2011NatPhys}.
For the considered gate figures, the best error reduction is obtained with the Uhrig and CPMG sequences: 
with $\omega_c= 5 \times 10^9$rad/s, they yield approximately two orders of magnitude improvement for a number of pulses $2n \sim 8-10$, until $\gamma_M \approx 10^8$s$^{-1}$. 
Such high-fidelity gate ($\epsilon \leq 10^{-6}$) could be obtained provided  
pulse rates $\Delta t^{-1}\approx 4 n \omega_c/\pi$ are available. 
One could think to operate with slower gates. However, according to our analysis, this 
has the drawback that decreasing $\omega_c$ would also increase the threshold value, $n_0 \propto \Omega/\omega_c$, 
where $\pi_y$-sequences give the minimum error. Therefore, in order to reduce the gate error
of two orders of magnitude a larger number of pulses would be required, resulting either in
an overall comparable pulse rate or in a gate lasting for longer time, thereby being more sensitive to 
Markovian decay channels. 
Operating with small pulse numbers is  desirable also in order to limit effects of pulse imperfections 
\cite{cywinsky2008PRB,Cywinski2014}.

\section{Conclusions}

In this paper we have investigated dynamical decoupling of local transverse $1/f$ noise in a two-qubit gate when the two
qubits operate at the optimal point. Our results for quasi-static noise are summarized in Table \ref{tab:scalings}. 
Decoupling of dynamic $1/f$ noise has been discussed in Section IV, based
on the analysis reported in Fig. \ref{fig:cutoff-analysis}. We have proven that two-qubit gate with errors 
$\sim 10^{-6}$, meeting the requirements for fault-tolerant quantum computation,
can be achieved by applying simultaneous, local pulses at proper times. This decouple while 
compute strategy 
applies under current experimental conditions of large-bandwidth $1/f$ noise.
Integrating gate operation with DD \cite{West2010PRL} guarantees the shortness
of the whole protocol avoiding further decoherence effects from Markovian noise.
Sequences we consider are simple and available
in many laboratories. In our scheme no encoding in a larger Hilbert space is
required, implying no resource overhead. We considered realistic $1/f$ spectra measured in superconducting qubits and  
investigated the stability of the protocols with UV (soft) cut-off. We found that, depending on the noise characteristics, the 
best dynamical scheme requires optimization of the trade-off between advantages of operating with 
larger qubit's coupling $\omega_c$ and anti-Zeno behavior.

In our analysis we assumed unbounded control by dc-pulses shined on each qubit simultaneously. 
This is necessary to preserve their interaction while averaging out their individual sources of defocussing.
Possible issues are the
sensitivity of the considered protocols to pulse timing and to what extent the analysis applies
when bounded control is considered. Pulse imperfections, like stochastic 
fluctuations of pulse sequence parameters, including duration and strength, and leakage to higher-energy states 
should be considered as well.
Addressing all these issues is beyond the scope of this paper. 
However, most of these questions have already emerged in similar contexts. Based on the existing literature, we can infer the following expectations and outlook. 

Robustness against imperfections and protection against experimental control errors in the DD pulses can be achieved
by robust optimal control~\cite{Montangero2007,Whaley2012}, concatenated DD sequences (CDD)  \cite{Khodjasteh2005,West2010PRL,Piltz2013}, composite pulses (CP)\cite{Cummins2003,Brown2004,Souza2012,Wang2014,Kabytayev2014}, 
dynamically corrected gates (DCG) \cite{Khodjasteh2009a,Khodjasteh2012} and  concatenated DCG \cite{Khodjasteh2010}. 
For bounded-strength non-Markovian environments, CDD has been shown to be advantageous with respect to PDD for state stabilization \cite{Khodjasteh2005}, for single qubit rotations \cite{Khodjasteh2007} and controlled-phase gates~\cite{West2010PRL}.
Recently, concatenated CP and DCG vastly improved the performance of a target single qubit gate
both in case of dephasing (amplitude) and transverse (detuning) noise and realistic $1/f$ spectra \cite{Kabytayev2014}.
A two-qubit controlled-NOT gate using CDD robust against imperfections of DD pulses has been realized
with trapped atomic ions coupled by an effective spin-spin interaction~\cite{Piltz2013}.
One could envisage that, in the schemes we considered, 
proper local CDD sequences or design of dynamically corrected two-qubit gates subject to
local transverse $1/f$  noise may result in efficient protocols also stable against imperfections. 

Pulse shaping offers the possibility to counteract the effect of noise during the finite duration $\tau_p$ of realistic pulses. 
By fulfilling simple analytic integrals, pulses can be shaped such that they approximate ideal, instantaneous pulses \cite{Pasini2008}. 
These optimized $\pi$ pulses are an excellent starting point for optimized dynamic decoupling schemes.  
Optimization of control of a single qubit gate in the presence of random telegraph noise \cite{Mottonen2006}, $1/f$ noise \cite{Whaley2008}
and interaction with a spurious quantum two-level system \cite{Rebentrost2009} has shown that realistic $\pi$ pulses can be made robust both to implementation errors and environmental noise. 
These works imply that treating realistic pulses as instantaneous is a reasonable 
approximation for $\tau_p$ smaller than the minimal interval between pulses of a given sequence. 
We therefore expect that our analysis applies to realistic, bounded control, provided the above conditions are satisfied.
Nevertheless, numerical optimization of control sequences devised to achieve entangling gates is a promising perspective. 
A high-fidelity CNOT-gate of inductively coupled flux qubits by quantum optimal control theory, considering leakage outside the computational space and Markovian decoherence, has been demonstrated in  \textcite{Huang2014}. 
Gate errors are found $\sim 10^{-5}$ considering both qubits at their optimal point and disregarding $1/f$ noise.
Our work provides complementary indications that suitably tailored decoupling strategies can efficiently decouple
the most detrimental noise sources. As a difference with quantum optimal control, the sequences considered in our work do not require
high-precision characterization of the parameters entering the Hamiltonian model. 
Remarkable improvements in gate fidelities have been recently obtained by optimal control in experiments with hybrid systems.
 In particular, numerically optimized control has been employed to steer the quantum evolution of a hybrid qubit system 
realizing a two-qubit gate \cite{Liu2013},
and integration of DD with quantum gates in an electron-nuclear spin register has  been proved \cite{VanDerSar2012}.
Using simultaneous DD sequences, the first capacitively coupled spin-qubits gate has been recently obtained \cite{Shulman2012} and
optimized control protocols able to improve the gate fidelity have been demonstrated \cite{Wang2015}. 
Should these tools be extended to other solid-state nanoscale devices, like superconducting nanocircuits, further advantages
and improved stability can be expected. 

\begin{acknowledgments}
This work has been partially 
supported 
by MIUR through Grant No. PON02-00355-3391233,
Tecnologie per l'ENERGia e l'Efficienza energETICa - ENERGETIC.
\end{acknowledgments}


\end{document}